\documentclass[useAMS,usenatbib,usegraphicx,useamssymb]{mn2e}
\usepackage{graphicx}
\usepackage{amssymb}
\usepackage{floatpag}
\usepackage{float}
\usepackage[section]{placeins}
\usepackage{relsize}
\usepackage{amsmath}
\usepackage{url}
\usepackage{longtable}
\usepackage{xcolor}
\usepackage{color}
\usepackage{supertabular}
\def \apj {ApJ}
\def \apjs {ApJS}
\def \apjl {ApJL}
\def \mnras {MNRAS}
\def \aap {A\&A}

\def \araa {ARAA}

\def \nar {New. A. Rev.}

\def \jcap {JCAP}

\def \col {black}
\usepackage{amsmath} % or simply amstext

\title[]{Constraining the rate and luminosity function of \emph{Swift} gamma-ray bursts}
\author[]{E. J. Howell$^{1}$\thanks{E-mail:eric.howell@uwa.edu.au}, D. M. Coward$^{1}$, G. Stratta$^{2}$, B. Gendre$^{3}$ and H.  Zhou$^{4}$\\
$^{1}$School of Physics, University of Western Australia, Crawley WA 6009, Australia\\
$^{2}$INAF - Osservatorio Astronomico di Roma, Via Frascati 33, I-00040 Monteporzio Catone (Roma), Italy\\
$^{3}$ARTEMIS, Observatoire de la C\^ote d’Azur, Boulevard de l’Observatoire B.P. 4229 F-06304 NICE Cedex 4, FRANCE\\
$^{4}$Department of Statistics and Finance, University of Science and Technology of China, Hefei, 230026, China
}

\begin{document}
\linespread{1.0}

\maketitle

\begin{abstract}
We compute the intrinsic isotropic peak luminosity function (LF) and formation rate of long gamma-ray bursts (LGRBs) using a novel approach. We complement a standard log\,$N$\,--\,log\,$P$ brightness distribution and $V_{\mathrm{max}}$ estimations with two observation-time relations: a redshift--observation-time relation (log\,$z$\,--\,log\,$T$) and a new luminosity--observation-time relation (log\,$L$\,--\,log\,$T$).
We show that this approach reduces degeneracies that exist between the rate and LF of a brightness distribution. To account for the complex triggering algorithm employed by \emph{Swift} we use recent results of \citet{Lien_2014ApJ} to produce a suite of efficiency functions. Using these functions with the above methods, we show that a log\,$L$\,--\,log\,$T$ method can provide good constraints on the form of the LF, particularly the high end. Using a sample of 175 peak luminosities determined from redshifts with well defined selection criteria our results suggest that LGRBs occur at a local rate (without beaming corrections) of $[\,0.7 < \rho_{0} < 0.8\,]\,\mathrm{Gpc}^{-3}\mathrm{yr}^{-1}$. Within this range, assuming a broken-power-law LF, we find best estimates for the low and high energy indices of $-0.95 \pm 0.09$ and $-2.59 \pm0.93$ respectively, separated by a break luminosity $0.80 \pm0.43 \times 10^{52}$\,erg\,s$^{-1}$.
\end{abstract}


%\graphicspath{{C:/Users/ejhowell/Documents/ASTRO/MATLAB/LogLLogT/for_paper/April2013/}{./}}

\section{Introduction}

Multi-wavelength observations of $\gamma$-ray bursts (GRBs) during the \emph{Swift} era have unambiguously confirmed these events to be the most luminous\footnote{In terms of electromagnetic radiation per unit solid angle.} and distant transients in the Universe \citep{Greiner_furthestGRB6pt7_08,2011arXiv1105.4915C}. A key objective of the \emph{Swift} mission was to obtain an accurate determination of the GRB luminosity function (LF) through the accumulation of redshift measurements \citep{Gehrels_Swift_2004}. Although \emph{Swift} has obtained over 200 redshifts, for long duration $\gamma$-ray bursts (LGRBs), the data is still insufficient to determine the LF accurately; additionally, the redshift distribution has been plagued by various selection effects \citep{Fiore2007A&A,coward_GRBDessert_08,Coward2012MNRASa}. These biases must be fully understood to gain an accurate representation of the intrinsic LF\footnote{The luminosity distribution of bursts irrespective of detection.}.

To circumvent these obstacles, many authors have chosen to employ the more abundant high energy data i.e. the brightness distribution of bursts \citep{horack94,Meszaros_ApJ_1996ApJ,sethi01, guetta05,Guetta2007JCAP,salvaterra_07,Salvaterra2010,Cao_2011MNRAS,Salvaterra2012ApJ}. The brightness distribution or log\,$N$--\,log\,$P$ distribution is a convolution of the source rate evolution and the intrinsic LF. Although this method is sensitive to the form of the LF, an obstacle often encountered is that mixing of the LF and rate evolution can introduce a degeneracy \citep{Firmani_04,Guetta2007JCAP}. This can be further complicated if one considers additional factors such as redshift evolution of the cosmic metallicity dependence or an evolving LF.

In this study we demonstrate a novel approach to this problem by complementing a standard log\,$N$--\,log\,$P$ analysis with two observation-time relations: Firstly, a new peak luminosity -- observation-time relation (log\,$L$--\,log\,$T$) will be used to scrutinise the estimated parameters of a log\,$N$--\,log\,$P$ distribution. Secondly, a redshift -- observation time relation (log\,$z$--\,log\,$T$) will be used to confirm limits on the range of possible values of the local rate density of LGRBs (without correction for beaming) obtained through a $V_{\mathrm{max}}$ estimation. An important part of our analysis will be an accurate representation of the \emph{Swift} triggering threshold. To do this we use recent results from \citet[][; L14 hereafter]{Lien_2014ApJ} to produce functions which are used to approximate the efficiency of \emph{Swift} triggering and the probability of observing a burst at a given redshift. We will show that these methods can provide both useful constraints on the LF of LGRBs. To perform our analysis, we will construct a sample of 175 bursts with redshifts confirmed through absorbtion spectroscopy and photometry and luminosity estimates calculated and corrected using the spectral parameters of \citet{Butler2007ApJ,Butler_2010}. The whole catalogue will also be made available online.

The paper will be organised as follows: In Section \ref{section_time_dimension} we will introduce the concept of using the observation time dependence of transients and will briefly describe some of the works which have successfully exploited this parameter. In Section \ref{section_swift_detection efficiency} we will discuss the results of L14 and present the efficiency functions that will be used in the analysis. In Section \ref{section_theoretical_framework}, we will set the theoretical framework for this study.

In Section \ref{section_logLlogT_theory} a new log $L$\,--log $T$ relation will be derived and in section \ref{section_logZlogT_theory} we will describe the log $z$\,--log $T$ relation. Section \ref{The logLlogT relation} will describe how the log $L$\,--log $T$ relation can indicate the form of the LF. After describing our data sample in section \ref{section_LGRB_data_sample} we will introduce the data extraction methods used for the observation time relations in section \ref{section_data_extraction}. We will determine an estimate of the local rate density of LGRBs in section \ref{section_rate_results}, after which we will probe the most likely parameters of the LF in section \ref{section_LF_results}. We  will conclude by summarising our findings in Section \ref{section_conclusions}.

\section{Exploiting the time dimension of GRBs}
\label{section_time_dimension}

\citet{howell_2013} illustrated how the time-record of GRB observations could be used as a tool to untangle different GRB populations. They did this by considering the \emph{rarest} events in a population i.e. those events from the tail of the distribution which occur at low-$z$ or have exceptional brightness in comparison with the average. The methodology is based on the use of extreme value statistics \citep{epstein_66} and follows the study of  \citep{coward_PEH_05,coward05b} who showed that these rare events impose a unique rate dependent statistical signature that can be described by the `probability event horizon' (PEH) concept. By recording successively rarer events as a function of observation-time, a data set -- termed \emph{PEH data} -- can be produced and constrained by a rate dependent model for peak flux -- log\,$P$--\,log\,$T$ \citep{howell07} or redshift -- log\,$z$--\,log\,$T$ \citep{howell_2013}.

\begin{figure}
  \includegraphics[scale = 0.52,bbllx = 0cm,bblly = 0.3cm, bburx = 13.2cm, bbury = 11cm,origin=lr]{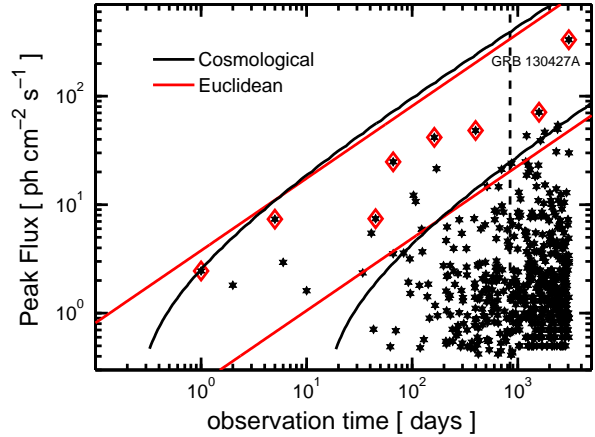}\\
  \caption{The \emph{Swift} LGRB peak flux data from April 2005 to June 2013 plotted against observation-time. The plot illustrates how the probability of observing a bright event increases with observation-time, $T$. Successively brighter events - termed \emph{PEH data} -  are indicated by diamonds. A 90\% confidence band constrains the data - this corresponds to the probabilities $\mathcal{P}=5\%$ (top) and $\mathcal{P}=95\%$ (bottom) of detecting at least one event within $T$. We see that the bright burst GRB 130427A is consistent with the prediction made in \citet{howell07} (indicated by the vertical dashed line). }
  \label{figure_swift_logPlogT_data}
\end{figure}

The basic concept is illustrated in Figure \ref{figure_swift_logPlogT_data} which shows the \emph{Swift} LGRB peak flux data plotted up to June 2013 against observation-time. Using a log-log plot it is apparent that successively brighter events have an observation-time dependence -- the longer you observe, the greater the probability of observing an exceptionally bright event such as GRB 130427A. Successively brighter events are indicated by diamonds -- these are the PEH data. The figure shows how PEH data is constrained using the Euclidean and cosmological log\,$P$--\,log\,$T$ models of \citep{howell07_APJL,howell_2013}. One should note that the data is not always as well behaved -- for example, there is always a possibility of a bright event occurring early in a time series. Therefore data extraction methods, which will be discussed later in Section \ref{section_data_extraction}, are employed to gain an optimal data set.

Adopting a time dependence allows the method to be used as a tool to predict the likelihood of future events. This is clearly illustrated in Figure \ref{figure_swift_logPlogT_data}. The vertical dashed line indicates the year 2007 when this relation was first published in \citet{howell07} using a smaller sample of \emph{Swift} bursts. We see that since this initial result two additional bursts consistent with the prediction have been added to the PEH sample. These include the recent bright burst GRB 130427A \citep{Perley_2013,Levan_130427A_2013}.

In this paper we will show that the log\,$z$--\,log\,$T$ relation can complement the more frequently used number count relations. As shown in \citet{howell_2013}, as PEH events approach the local low-$z$ regime rapidly, the GRB selection function \citep{Coward2007NewAR} and high-$z$ selection effects such as the `redshift desert' \citep{coward_GRBDessert_08,Coward2013MOR} have a negligible effect. Therefore, observation time relations can be used as a test of parameter compatibility without  consideration of selection biases. To examine the LF of LGRBs, section \ref{section_logLlogT_theory} will extend previous studies by introducing a new cosmological log\,$L$--\,log\,$T$ relation.

\section{The detection efficiency of \emph{Swift}}
\label{section_swift_detection efficiency}

\textcolor{\col}{
Modeling the triggering criteria of an astronomical instrument can be critical and an oversimplified approach can lead to errors in the determination of population parameters or incorrect assumptions of the completeness of a sample.}
\textcolor{\col}{
For instruments prior to \emph{Swift}, such as BATSE, assuming a single detection threshold based on an increased photon count rate above background was a reasonable approximation \citep[however, see][for further discussion of BATSE]{Shahmoradi2011MNRAS}. However, for \emph{Swift} a highly complex triggering algorithm has been adopted based on 674 different trigger criteria (see L14 for a comprehensive description).}

\textcolor{\col}{
Approximating the triggering response of such a sophisticated instrument as \emph{Swift} is highly challenging. Numerous studies have approached this problem by assuming a single detection threshold based increased photon count rate above background \citep[][]{Le2007ApJ,Guetta2007JCAP,Elliott2012,Salvaterra2012ApJ} (the sole triggering criterion used for BATSE), have used analytical approximations of the \emph{Swift} triggering efficiency \citep{Qin2010,Wanderman2010,Lu_2012ApJ} or have adopted an effective threshold based on the luminosity-redshift distribution \citep{Kistler2008ApJ,Cao_2011,Kistler2013}.}

In this study we use an alternative approach based on the recent comprehensive study of L14, who have mimicked the 674 criteria of the \emph{Swift} triggering algorithm through a Monte Carlo approach. This included monitoring increased count rates on different timescales, energy bands and regions of the focal plane, periods of foreground (periods of strong emission) and background periods. Using BAT detected GRB light curves with redshifts, a mock rest frame sample was created and converted to photon counts corresponding to different incidence angles. Through a simulation of 50000 bursts, a detailed comparison of the triggered sample with the observed \emph{Swift} distributions enabled both a thorough interrogation of the \emph{Swift} detector response and a determination of the global parameters of the LGRB population.

To produce a set of efficiency functions for the \emph{Swift} instrument we make use of the resulting L14 distributions (simulated and detected) in peak flux and redshift. This is a different approach to other studies which have determined similar functions based solely on the \emph{Swift} detected distributions \citep{Wanderman2010} or have used a simple scaling criteria \citep{howell_2013}. The adoption of these efficiency functions is important in this study. For example, the determination of a local rate density using the $V_{\mathrm{max}}$ method (see later section \ref{section_rate_results}) uses only a small sample of the closest occurring bursts -- in this scenario, the flux threshold of the detector has a highly significant bearing on the final estimate. The remainder of this section will present and describe the analytical forms of these functions.
%}

\subsection{The \emph{Swift} peak flux efficiency function}
\label{peak_flux_efficiency}
\begin{figure}
\includegraphics[scale = 0.52,bbllx = -0.5cm,bblly = -0.2cm, bburx =9cm, bbury = 11.5cm,origin=lr]{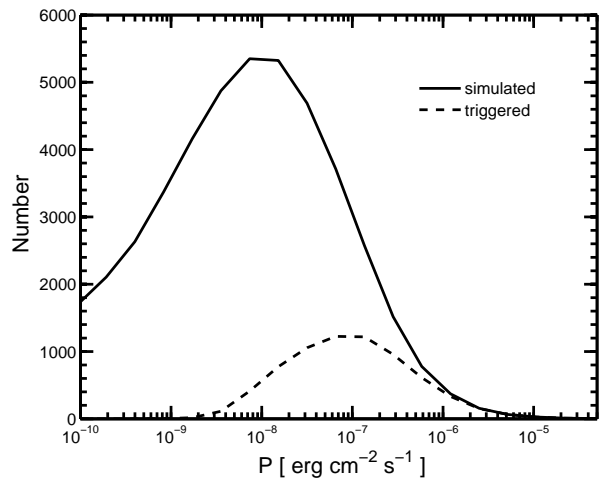}\\
  \caption{The simulated and triggered peak flux sample of L14 based on a comprehensive study of the \emph{Swift} detection efficiency. The broad simulated distribution of peak fluxes (50,000) is able to sample the detection response across a wide range of values.}
  \label{figure_swiwt sim_sample}
\end{figure}

%\textcolor{\col}{
Figure \ref{figure_swiwt sim_sample} shows the simulated and triggered peak flux sample of L14. The triggered sample is produced through a comprehensive reproduction of the \emph{Swift} triggering algorithm. Although the simulated data was produced using a specific set of model parameters, we note the broad simulated distribution of peak fluxes (50,000) samples the detection response across a significantly wider range of values than generally considered. In comparison to the other techniques to model the \emph{Swift} triggering algorithm, we suggest that the triggered sample (obtained through careful modeling of all 674 criteria) provides an adequate representation of the \emph{Swift} response for use in this study.

\begin{figure}
\includegraphics[scale = 0.51,bbllx = 0.0cm,bblly = 0.0cm, bburx =9cm, bbury = 12cm,origin=lr]{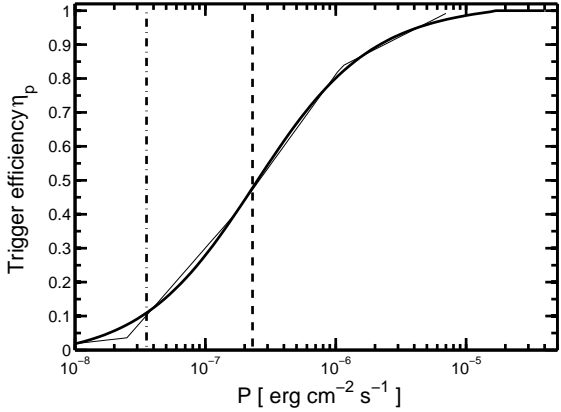}\\
  \caption{\color{\col}The triggering efficiency function of \emph{Swift} as a function of peak flux for LGRBs adapted from the simulated sample of L14. The efficiency of the \emph{Swift} trigger is shown by the thin line. Two Heaviside step function approximations are shown for comparison at values of 0.4 ph cm$^{-2}$ s$^{-1}$ and 2.6 ph cm$^{-2}$ s$^{-1}$.  }
  \label{figure_flux_eff}
\end{figure}

By scaling of the simulated peak flux and triggered distributions of L14, we obtain the following trigger efficiency function:
%}
\begin{equation}\label{swift_pf_efficiency_curve}\color{\col}
 \eta_{P}(P) =  \frac{  a(b + cP/P_{0})  }{ ( 1 + P/d\,P_{0})   }\,.
\end{equation}
\textcolor{\col}{
\noindent Within the range $5.87\times 10^{-9}$ $ < P < $ $1.69\times 10^{-5}$ $\eta_{P}$  the function takes the parameters: \{a = 0.47,; b = -0.05\,; c = 1.46\,; d = 1.45\,  $P_{0}= 1.6\times10^{-7}$\}; below and above this range the function equals 0 and 1 respectively. The form of $\eta_{P}(P)$ is shown in Figure \ref{figure_flux_eff}. We apply this function to the simulated peak flux population within the above range and are able to reproduce the triggered population of L14 with a statistical computability of $P_{\mathrm{KS}}>99\%$ as measured by a two sample Kolmogorov-Smirnov test.}

\textcolor{\col}{
As discussed earlier, many studies use a simple Heaviside step function to account for the triggering threshold of \emph{Swift}.
For illustration Figure \ref{figure_flux_eff} compares $\eta_{P}(P)$ with two step function approximations: a value of 0.4 ph cm$^{-2}$ s$^{-1}$ similar to the value often adopted for BATSE; a value of 2.6 ph cm$^{-2}$ s$^{-1}$ used by \citet{Salvaterra2012ApJ} to produce a sample of bursts with a completeness of 95\%. We see that even the higher of these two thresholds indicates a triggering efficiency of no more than 50\%. It is apparent how the adoption of such approximations, or similar estimates based on an effective detection threshold determined from the luminosity-redshift plane, could be problematic when estimating the relative contributions of bright and dim bursts.}

\subsection{The \emph{Swift} redshift efficiency function}
\label{redshift_efficiency}

To model the efficiency as a function of redshift, \emph{z}, we use the result of Fig. 16a of L14 to produce the following piecewise analytical fit:

\begin{equation}\label{equation_z_eff}
   \eta_{z}(z)=\biggl \lbrace{ \begin{array}{lll}
                   a+b\, \mathrm{exp}(-z/c) & z\,<\,5.96
                   \\
                   0.02 & z\,>\,5.96  \\

                 \end{array}}
\end{equation}

\noindent which takes the values \{a = -0.01; b = 1.02; c = 1.68\}. We note that this function provides the fraction of bursts at each redshift interval that have passed the peak flux triggering criteria; therefore, if working in solely redshift space -- such as when using the log\,$z$--\,log $T$ distribution in section \ref{section_rate_results} -- one should also scale the rate with the average peak flux efficiency of 14\% determined by L14.
\begin{figure}
\includegraphics[scale = 0.52,bbllx = 0cm,bblly = 0.5cm, bburx =9cm, bbury = 10cm,origin=lr]{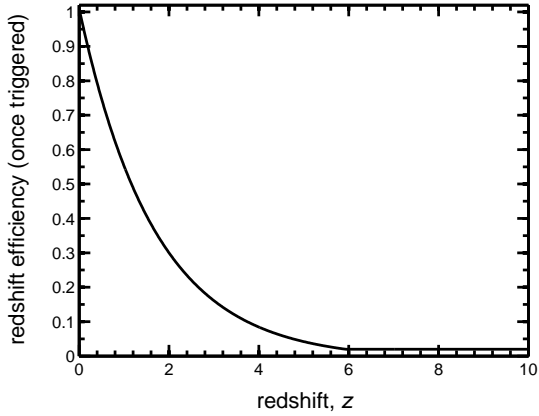}\\
  \caption{The redshift efficiency curve for Swift adapted from Fig 16a of L14. The figure represents the detection efficiency as a function of redshift.}
  \label{figure_redshift_eff}
\end{figure}

We will apply these functions in our analysis of the \emph{Swift} data sample -- it will be interesting to draw comparisons between the global parameters determined in this study with those determined through simulation of the \emph{Swift} response by L14.

\section{Theoretical Framework}
\label{section_theoretical_framework}

\subsection{GRB Flux and luminosity relations}
\label{GRB Flux and luminosity relations}

An isotropic equivalent luminosity in the source frame (erg s$^{-1}$) can be calculated from:

\begin{equation}\label{eq_rest_frame_luminosity}
    L = 4 \pi d_{\mathrm{L}}(z)^{2} P \frac{k(z)}{b}
\end{equation}

\noindent where $P$ is the energy flux (erg s$^{-1}$ cm$^{-2}$) in the observed energy band $E_{\mathrm{min}}\hspace{-0.5mm}<\hspace{-0.5mm}E\hspace{-0.5mm}<\hspace{-0.5mm}E_{\mathrm{max}}$, $d_{\mathrm{L}}(z)$ is the luminosity distance and $k(z)$ and $b$ are correction terms to convert the observed flux in the detector band (for \emph{Swift} this is 15--150 keV) to the rest frame band 1--$10^{4}$ keV . The first of these correction terms is the bolometric correction $b$ which accounts for the differing fraction of gamma ray energy seen in the detector band \citep{Imerito_08,Wanderman2010}:

\begin{equation}\label{eq_bol_correction}
b = \int^{E_{\mathrm{max}}}_{E_{\mathrm{min}}}E S(E)dE / \int^{10000}_{1}E S(E)dE\,,
\end{equation}

\noindent where $S(E)$ is the rest-frame photon spectrum (ph cm$^{-1}$ keV$^{-1}$) multiplied by a factor of $E$ (energy in keV) as part of the conversion to energy units. The term $k(z)$ the cosmological $k$-correction and given by:

\begin{equation}\label{eq_kcorrection}
k(z) =  \int^{E_{\mathrm{max}}}_{E_{\mathrm{min}}}E S(E)dE / \int^{E_{\mathrm{max}}(1 + z)}_{E_{\mathrm{min}}(1 + z)}E S(E)dE\,.
\end{equation}

\noindent Rearranging and substituting for $b$ and $k(z)$ in the above equation yields the familiar relation for energy flux\footnote{This relation is valid for a peak flux with energy units (erg\,s$^{-1}$\,cm$^{-2}$). If using a peak photon flux (ph\,s$^{-1}$\,cm$^{-2}$) there is an additional factor of $(1 + z)$.}:

\vspace{-0.5mm}
\begin{equation}\label{eq_peakflux}
    P = \frac{ \int^ {(1 + z)E_{\mathrm{max}}}_{(1 + z)E_{\mathrm{min}}} E S(E)\mathrm{d}E}{4\pi d_{\mathrm{L}}(z)^{2}}\,.
\end{equation}

\noindent For long duration GRBs the function $S(E)$ is typically modeled by a Band function \citep{Band03} which we use with high and low energy spectral indices of -2.25 and -1 and a peak energy of 511 keV (in the source frame). Unless spectral forms and parameters are available (such as those used in determining the luminosity corrections in section \ref{section_LGRB_data_sample}) we will assume these values.

\subsection{GRB Luminosity Function}

To model the LGRB Luminosity Function (LF), we use a Broken Power law model (BPL) model which takes the form:

\begin{equation}
\Phi(L) \propto
\biggl \lbrace{
\begin{array}{ll}
\hspace{1.0mm}\left(L/L_{*}\right) ^{\alpha} \hspace{3.0mm} L < L^{*}\\
\hspace{1.0mm}\left(L/L_{*}\right) ^{\beta}  \hspace{3.0mm} L \geq L^{*}\\
\end{array} }
\end{equation}

\noindent with, $L$ is the isotropic rest frame luminosity in the 1-10000\,keV energy range and $L_{*}$ a characteristic cutoff scaling that separates the two slopes $\alpha$ and $\beta$. The additional power law in comparison with single power law forms LF allows to examine the low and high luminosity parts of the distribution. We follow the studies of \citet{Meszaros_ApJ_1996ApJ, Meszaros_ApJ_1995ApJ, Reichart_ApJ_1997,Butler_2010} and assume no luminosity evolution with redshift.

\subsection{GRB source rate evolution}

\textcolor{\col}{To obtain a source rate evolution model for LGRBs with redshift, $R_{\rm GRB}(z)$, we use the piecewise function of \citet{Wanderman2010}:}

\begin{equation}\label{eq_sfr}
\color{\col}
   R_{\mathrm{GRB}}(z=0)=\biggl \lbrace{ \begin{array}{lll}
                   ( 1 + z)^{a} & z\,<\,z_{*}
                   \\
                   ( 1 + z_{*})^{a - b}(1 + z)^{b} & z\,>\,z_{*}
                    \\
                 \end{array}}
\end{equation}
\noindent
\textcolor{\col}{
with values of $z_{*}=3.6$, a=2.1 and b=-0.7 based on the recent study of W14.
}

\subsection{The all sky event rate equation of GRBs}

The number of GRBs per unit time within the redshift shell $z$ to $z + \mathrm{d}z$ with luminosity $L$ to $L + \mathrm{d}L$ is given by:
\vspace{-0.5mm}
\begin{equation}
\label{dN}
\frac{\mathrm{d}N}{\mathrm{d}t \mathrm{d}z \mathrm{d}L }\hspace{0.5mm} = \psi(z)
 \frac{\mathrm{d}V (z) }{\mathrm{d}z } \frac{ R_{\mathrm{GRB}}(z) }{( 1 + z)}\,\mathrm{d}z{\hspace{0.5mm}}\Phi(L)\,.
\vspace{0mm}
\end{equation}

\noindent Here the $(1 + z)$ factor accounts for the time dilation of the observed rate by cosmic expansion; its inclusion converts source-count information to an event rate. The co-moving volume element:
\vspace{-0.5mm}
\begin{equation}\label{dvdz}
\frac{\mathrm{d}V}{\mathrm{d}z}= \frac{4\pi c}{H_{0}}\frac{d_\mathrm{\hspace{0.25mm}L}^{\hspace{1.5mm}2}(z)}{(1 +
z)^{\hspace{0.25mm}2}\hspace{0.5mm}h(z)}\,,
\end{equation}

\vspace{-0.5mm}
\noindent describes how the number densities of non-evolving objects locked into Hubble flow are constant with redshift. The quantity $h(z)$, is the normalized Hubble parameter,
\vspace{-0.5mm}
\begin{equation}\label{hz}
h(z)\equiv H(z)/H_0 = \big[\Omega_{\mathrm m} (1+z)^3+ \Omega_{\mathrm \Lambda} \big]^{1/2}\,,
\end{equation}

\noindent where $\Omega_{\mathrm m} + \Omega_{\mathrm \Lambda}=1$ \citep[for further details see][]{Carroll_ARAA_1992}. For a `flat-$\Lambda$' cosmology, we employ the most recent cosmological parameters measured by Planck of $\Omega_{\mathrm
m}=0.32$, $\Omega_{\mathrm \Lambda}=0.68$ and
\mbox{$H_{0}=67$ km s$^{-1}$ Mpc$^{-1}$} for the Hubble parameter at the present epoch \citep{Planck2013}.

\section{The log\,$L$--log\,$T$ relation}
\label{section_logLlogT_theory}

From equation \ref{dN}, the rate of GRBs with a peak luminosity greater than $L$ observed by an instrument with sky coverage $\Omega$ is given by:
\vspace{-0.5mm}
\begin{equation}
\label{eq_integral_lum}
\hspace{-1mm}
\dot{N}(>L)=\frac{\Omega}{4 \pi} \hspace{-4mm}\int \limits_{\hspace{4mm}L_{\mathrm{min}(z)} }^{\hspace{4mm}\infty } \hspace{-4mm} \Phi(L) \mathrm{d}L \hspace{-3mm}\int \limits_{\hspace{1mm}0}^{\hspace{4mm}\infty}
\hspace{-1mm}\frac{\mathrm{d}V (z) }{\mathrm{d}z } \frac{ R_{\mathrm{GRB}}(z) }{( 1 + z)}\,\mathrm{d}z \,,
\vspace{0mm}
\end{equation}

\noindent were $L_{\mathrm{min}}(z)=\mathrm{max}[L,L(p_{\mathrm{min}},z))]$: here $L(P_{\mathrm{min}},z)$ is obtained through equation \ref{eq_peakflux} and is the minimum luminosity required by a burst at redshift $z$ to produce a peak flux of $P_{\mathrm{min}}$. This equation was used by \citet{Salvaterra2010} to estimate the number of bursts with peak luminosity $> 10^{53}$ in the \emph{Swift} sample.

To introduce an observation-time dependence, $T$, we follow the probability event horizon concept of \citet{coward_PEH_05} and note that as GRBs are independent of each other, their observation-times will follow a Poisson distribution in time. Therefore, the temporal separation between events will follow an exponential distribution defined by a mean number of events, $\dot{N}(> L)\,T$. The probability $\mathcal{P}$ for at least one event $> L$ is given by:
\vspace{-0.5mm}
\begin{equation}\label{eq_Lpeh}
\mathcal{P}(n \ge 1;\dot{N}(> L),T)=  1 - e^{-\dot{N}(> L) T} = \mathcal{P} \,.
\end{equation}

\noindent For this equation to remain satisfied with increasing observation-time:
\vspace{-0.5mm}
\begin{equation}\label{eq_logLlogT}
\dot{N}(> L) T =  |\mathrm{ln}(1 - \mathcal{P})| \,.
\end{equation}

\noindent Equating the above equation for $L$ and $T$ we obtain a relation for the evolution of isotropic luminosity as a function of observation-time. By setting $\mathcal{P}$ to some arbitrary value, log\,$L$--\,log $T$ curves can be obtained numerically through equations \ref{eq_integral_lum} and \ref{eq_logLlogT}. Following \citet{howell_2013,howell07_APJL,howell07} one can plot upper and lower thresholds by setting $\mathcal{P}=(0.95;0.05)$ -- we will refer to these curves as 90\% PEH bands.
\begin{figure}
\includegraphics[scale = 0.45,bbllx = -2.0cm,bblly = 0.5cm, bburx =9cm, bbury =14cm,origin=lr]{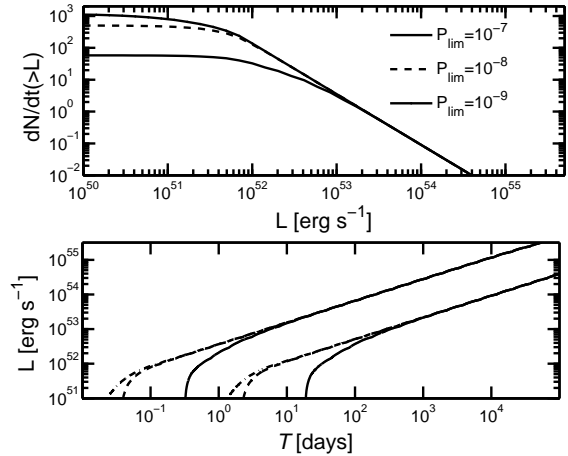}
  \caption{The functions $\dot{N}(> L)$ (top panel) and log\,$L$--\,log $T$ (lower panel) for the same model parameters but using different values of $P_{\mathrm{min}}$. The flux limit has little effect on the $\dot{N}(> L)$ above $10^{53}$ erg s$^{-1}$ -- for the upper log\,$L$--\,log $T$ 90\% PEH threshold (which gives the 5\% peak luminosity probability within time $T$) this corresponds to around 4 days of observation time.}
  \label{figure_lumpeh_intlum}
\end{figure}

We note that equation \ref{eq_integral_lum} is dependent on an estimate of $P_{\mathrm{min}}$ for the lower limit of the integral over the LF. As shown in section \ref{section_swift_detection efficiency} it is difficult to model the \emph{Swift} detection efficiency through a single value of $P_{\mathrm{min}}$. We find however, that the log\,$L$--\,log $T$ circumvents this obstacle through its dependence on the most luminous events. Figure \ref{figure_lumpeh_intlum} illustrates this property by comparing the functions $\dot{N}(> L)$ (top panel) and log\,$L$--\,log $T$ (lower panel) models for the same parameters but using different values of $P_{\mathrm{min}}$. The flux limit has little effect on the $\dot{N}(> L)$ above \mbox{$10^{53}$ erg s$^{-1}$} -- for the upper log\,$L$--\,log $T$ 90\% PEH threshold this corresponds to around 4 days of observation time, equal to that of the first PEH data point used in this study( see later in table \ref{table_LPEH_data}). We find similar convergence of the log\,$L$--\,log $T$ curves for rates within 0.2-0.9 Gpc$^{-3}$yr$^{-1}$  We therefore use a value of $P_{\mathrm{min}}= 10^{-8}$ erg s$^{-1}$cm$^{-2}$ noting that the adoption of this value will not influence our results.

\section{The log\,$z$--\,log $T$ relation}
\label{section_logZlogT_theory}

One can extend the arguments of the previous section to derive a log\,$z$--\,log $T$ relation \citep{howell_2013}. From equation \ref{dN} the rate of GRBs observed by an instrument with sky coverage $\Omega$ within a redshift limit $z_{\rm{L}}$ is given by:
\begin{equation}
\label{eq_integral_z}
\vspace{6mm}
\hspace{-1mm}\dot{N}(< z_{\rm{L}})=  \hspace{-1mm}\frac{\Omega}{4 \pi} \hspace{-12mm}\int\limits_{\hspace{12mm}L_{\mathrm{min}}(\mathrm{P_{min}},z_{\mathrm{L}}) }^{\hspace{7mm}L_{\mathrm{max}} } \hspace{-12mm}\Phi(L)\,\mathrm{d}L \hspace{-2mm}
\int\limits_{0}^{\hspace{5mm}z_{\rm{L}}}
\hspace{0mm}\frac{\mathrm{d}V (z) }{\mathrm{d}z }  \frac{\eta_{z}(z)\, R_{\mathrm{GRB}}(z) }{( 1 + z)}\,\mathrm{d}z \,,
\vspace{-2mm}
\end{equation}
\noindent with $z_{\mathrm{L}}$ obtained by applying the value, $P_{\mathrm{min}}$, to equation \ref{eq_peakflux}; the quantity $\eta_{z}$ is the efficiency of obtaining a redshift (sub-section \ref{redshift_efficiency}).

A similar argument as that used to determine equation \ref{eq_logLlogT}, yields the following relation for the temporal evolution of redshift:
\vspace{-0.5mm}
\begin{equation}\label{eq_logZlogT}
\dot{N}(< Z_{\rm{L}}) T =  |\mathrm{ln}(1 - \mathcal{P})|. \\
\end{equation}

\noindent This equation can be equated for $T$ and $z$ to set a spatial dependence on GRB populations. Curves of log\,$z$--\,log $T$ for $\mathcal{P}=(0.95;0.05)$ can be obtained numerically through equations \ref{eq_integral_z} and \ref{eq_logZlogT}. As before, these will be referred to as 90\% PEH bands.

As shown above, the log $z$\,--\,\,log\,$T$ and log $L$\,--\,\,log\,$T$ relations are derived seamlessly from standard integral distributions. Thus, model parameters obtained by fitting to a differential brightness distribution should satisfy the two observation-time relations presented above (equations \ref{eq_logLlogT} and \ref{eq_logZlogT}).

An advantage of this technique is that the PEH sample in redshift space is predominantly from the closest events. Therefore consideration of high-$z$ selection bias is not essential.
We will use log\,$z$\,--\,log\,$T$ relation later in section \ref{section_rate_results} to validate our estimate of (beaming uncorrected) local rate density, $\rho_{0}$.

\section{The log\,$L$--log\,$T$ relation as a probe of the LGRB luminosity function}
\label{The logLlogT relation}
\begin{figure*}
        \includegraphics[scale = 0.48,bbllx = 700pt,bblly =10pt, bburx = 940 pt, bbury = 530 pt,origin=lr]{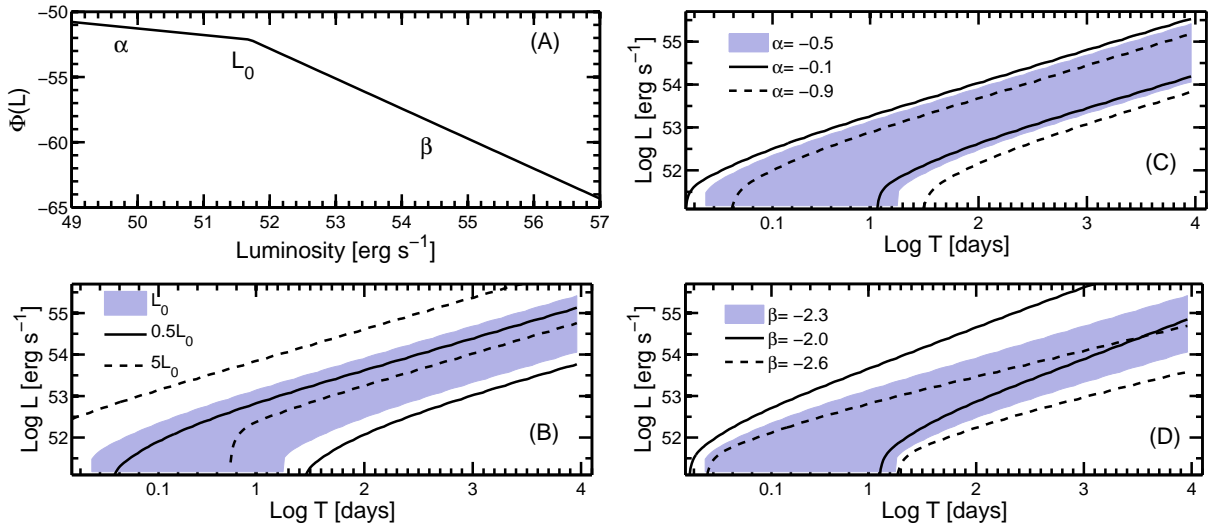}\\
  \caption{\textbf{Panel A} shows the broken power law LF model adopted for this study. For illustration we adopt the parameters $L_{*} =5\times 10^{51}\, \mathrm{erg} \mathrm{s}^{-1}, \alpha =-0.5$ and $\beta =-2.3$. The shaded portions of each of the remaining panels show the log\,$L$\,--\,log\,$T$ 90\% PEH bands corresponding to these parameters - the upper and lower curves in each case correspond with the probabilities given Figure \ref{figure_swift_logPlogT_data}. The solid and dashed lines in the panels B--D show how the curves are modified by changing each of the parameters: $L_{*}$, $\alpha$ and $\beta$. \textbf{Panel B:} increasing/decreasing the value of the break Luminosity, $L_{*}$, results in a vertical increase/decrease of the log $N$--log $L$ curve.  \textbf{Panel C:} increasing/descreasing the low end slope represented by $\alpha$ shifts the curves in the positive/negative horizontal direction. \textbf{Panel (D)} increasing/descreasing the value of the high end slope ($\beta$) produces a gradient change.}
  \label{fig_logLlogT_curves}
\end{figure*}

Figure \ref{fig_logLlogT_curves} shows how the log\,$L$--log\,$T$ 90\% PEH bands can indicate of the form of the LF. For illustration we adopt an arbitrary broken power law LF with parameters: $L_{*} =5\times 10^{51}\, \mathrm{erg} \mathrm{s}^{-1}, \alpha =-0.5$ and $\beta =-2.3$. This LF is shown in panel (A) and the corresponding log\,$L$--log\,$T$ curve is shown the other panels as a shaded component. Solid and dashed lines in each of the panels B-D illustrate how the log $L$\,--\,\,log $T$ curves are modified by changing one of the parameters of the LF. Below, we discuss and provide physical interpretations for these changes:
 \begin{itemize}
   \item Panel (B) shows that increasing/decreasing the value of the break Luminosity ($5\times L_{*}; 0.5\times L_{*}$) produces a vertical increase/decrease of the log $L$--log $T$ curves. A higher value of $L_{*}$ results in a greater probability of a bright event; thus an increased probability of a more energetic event at early observation times.
   \item Panel (C) shows that increasing/descreasing the gradient of the low end slope through $\alpha$ offsets the curves in the positive/negative horizontal direction. Increasing $\alpha$ produces a greater proportion of dimmer bursts - therefore, a lower probability and thus, an increased waiting time, for a high luminosity event.
   \item Panel (D) illustrates how increasing/descreasing the value of $\beta$ produces a gradient change. A flatter value of the bright-end slope produces a greater proportion of bright burst; this corresponds with an increase in the gradient of the probability curves.
 \end{itemize}
The sensitivity of the log\,$L$--log\,$T$ PEH curves to the parameters ($L_{*}$\, $\alpha$, $\beta$), means that estimates obtained through a brightness distribution (log\,$N$--log\,$P$) can be validated. In particular, employing this complementary method can help to untangle degeneracies that are encountered in using a log\,$N$--log\,$P$ fit. Before we conduct our analysis, in the next two sections we will describe the LGRB data sample and then discus how to extract a PEH data set.

\section{The LGRB Data Sample}
\label{section_LGRB_data_sample}
\subsection{The redshift data sample}
\label{subsection_redshift data}
In recent studies \citet{zhang_openQs_2011} and \citet{Bromberg2012} have suggested that the $T_{90}=2\,\mathrm{s}$ division of short and long GRBs based on the BATSE bimodial distribution \citep{Kouveliotou_1993} is a detector dependent categorisation and therefore not appropriate for \emph{Swift} bursts. Other studies have suggested an intermediate duration class of bursts between these two classes \citep{Horvath2010}. Additionally sub-luminous GRBs \citep[SL-GRBs;][]{howell_2013,Virgilii_LLGRBs_08,Imerito_08,Daigne_2007,cobb_06,coward_LLGRB_05}  and short GRBs with extended emissions \citep[SGRB-EEs;][]{norris_2011}) have been suggested to be members of sub-populations of burst.
To obtain a selection of \emph{Swift} LGRBs, rather than employing a $T_{90}$ cut, we use the catagorisations given in the Jochen Greiner online catalogue (JG) of well localized GRBs\,\footnote{\url{http://www.mpe.mpg.de/~jcg/grbgen.html}}. As the burst catagorisations and redshifts in this catalogue are subject to review through follow up studies we find it a useful resource to isolate a LGRB sample \citep[for example, the catalogue was recently updated using 15 new redshifts from the TOUGH survey][]{Hjorth_TOUGH_2012ApJ}.

We use data up to June 2013 which includes 232 redshifts of which 209 have secure redshifts (uncertain redshifts were omitted). To arrive at the redshift sample given above we have excluded three SL-GRBs, 060218, 060505 and 100316D (see \citet{howell_2013} for further discussion of these bursts), 7 bursts catagorised as SGRB-EEs (all but one have $T_{90}>2\,\mathrm{s}$) \citep{norris_2011} and 3 bursts (101225A, 111209A \& 121027A) strongly suggested to be part of an ultra-long GRB population \citep{Gendre_2012,levan_2013,Stratta2013}.

\begin{figure}
    \includegraphics[scale = 0.55,bbllx = 100pt,bblly =290pt, bburx = 500 pt, bbury = 600 pt,origin=lr]{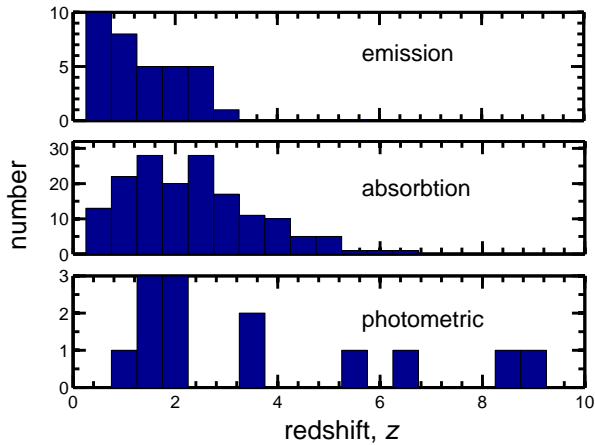}\\
  \caption{The redshift distributions of \emph{Swift} observed LGRBs up to June 2013 separated into data obtained though absorbtion and emission spectroscopy and photometrically. The photometric redshift distribution is approximately uniform  across the observed range. In comparison, the sample obtained through host galaxy emission spectra are only observed within a limited range $0.3 < z < 2.8$. }
  \label{GRB_Redshift_Dist}
\end{figure}

Figure \ref{GRB_Redshift_Dist} shows the sample of 209 bursts with redshifts determined through emission, absorbtion and photometry. For redshifts determined though multiple criteria, absorbtion takes precedence followed by photometry. The histograms show that photometrically determined bursts are detected almost uniformly across redshift, while host galaxy emission spectra cover a limited range $0.3 < z < 2.8$.
From our sample of 209 redshifts we select the 175 bursts obtained through absorbtion spectra (162 of the sample; or 78\%) or photometrically $(13; 6\%)$.  A two sample Kolmogorov-Smirnov test (KS) shows these two samples to be compatible at the KS probability $P_{\mathrm{KS}}=22\%$ level. We follow \citet{Wanderman2010} and omit the sample of redshifts obtained through host galaxy emission spectra which, due to their narrow redshift range, are not statistically compatible with the absorbtion sample $(P_{\mathrm{KS}} \sim 10^{-3}\%$).

\subsection{Luminosity data}
\label{subsection_luminosity data}
%\textbf{NOTE: How many estimated using the band function ?}\\
%\\
To calculate isotropic peak luminosity data, $L$, from the redshift sample we use equations \ref{eq_rest_frame_luminosity} -- \ref{eq_kcorrection}. Peak energy flux data is taken from Butler's online catalogue of \emph{Swift} BAT Integrated Spectral Parameters \footnote{\url{http://butler.lab.asu.edu/Swift/bat_spec_table.html}}. This catalogue, an extension of the work presented in \citet{Butler2007ApJ,Butler_2010}, has circumvented the nominal BAT upper energy of 150 keV to produce accurate values of $E_{\mathrm{iso}}$ through a novel Bayesian approach. Peak spectral energies from the BATSE catalogue have been used to set a strong prior on the range of GRB model parameters specified by one of three models of increasing complexity: a simple power law, a power law times an exponential cutoff and a Band function \citep{Band93}. The resulting spectra were shown to be in agreement with observations from satellites operating at much broader energy ranges e.g. \emph{Konus-Wind} (10--770 keV) and \emph{Suzuku} (0.3–-600 keV). We use the spectral parameters catalogued for each burst to calculate the $k$ and bolometric correction terms given in section \ref{GRB Flux and luminosity relations}.

\begin{figure}
\includegraphics[scale = 0.55,bbllx = 100pt,bblly =290pt, bburx = 500 pt, bbury = 600 pt,origin=lr]{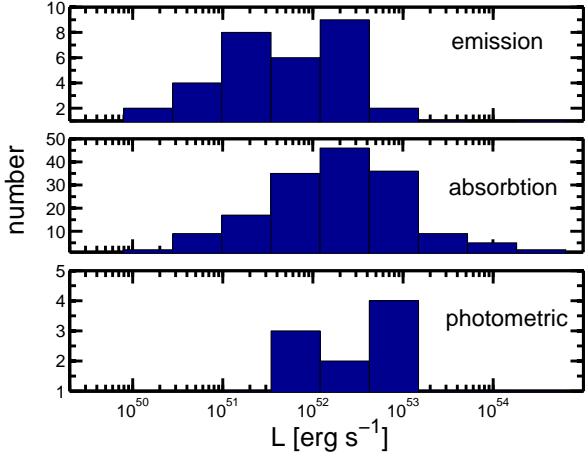}\\
  \caption{The peak luminosity distributions of \emph{Swift} observed LGRBs up to June 2013 obtained using absorbtion and emission spectroscopy as well as photometrically. The sample obtained through host galaxy emission spectra are clearly biased towards lower values of luminosity. }
  \label{GRB_Lum_Dist}
\end{figure}

Figure \ref{GRB_Lum_Dist} shows the sample of 209 burst luminosities using the redshifts determined through emission, absorbtion and photometry. As expected, the sample obtained through host galaxy emission spectra are clearly biased towards lower values of luminosity. This observation is confirmed through a KS test in which a value of $P_{\mathrm{KS}}\sim1\%$  is obtained between the absorbtion and emission samples. The photometric sample are compatible with the emission sample at a level of $P_{\mathrm{KS}}=23\%$ and are therefore included in our final sample of 175 bursts. Our full data sample is shown in Table \ref{table_lum_data} and is available online\footnote{Available at \url{http://www.ejhowell.com/data/}}.

\begin{figure}
  \includegraphics[scale = 0.55,bbllx = 1pt,bblly =1pt, bburx = 500 pt, bbury = 300 pt,origin=lr]{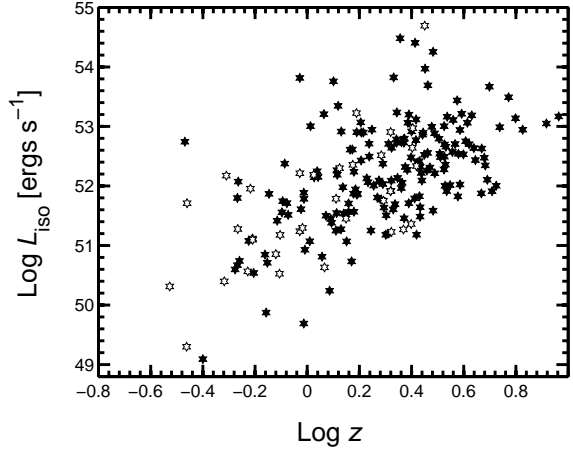}
  \caption{The luminosity--redshift distribution of \emph{Swift} LGRBs shown as filled black stars. The bursts with redshifts determined through emission spectra are shown by unfilled stars - these bursts are predominantly at lower-$z$.}
  \label{GRB_Redshift_Luminosity}
\end{figure}

Figure \ref{GRB_Redshift_Luminosity} shows the luminosity--redshift distribution (filled stars). The sample determined through emission spectra are also shown (unfilled stars). We test the statistical compatibility of our samples using a 2D-KS \citep{Peacock83,FF87} test in the $L_{\mathrm{iso}}$--$z$ plane. For the absorbtion/photometric samples we obtain
$(P_{\mathrm{KS,2d}} \sim 17\%)$ signifying good compatibility. The decision to omit emission data is further justified through a value of $(P_{\mathrm{KS,2d}} \sim 10^{-3}\%)$ for the absorbtion/emission data samples.

In the next section we will outline the extraction procedure used to obtain PEH data. We will then be well equipped to constrain this data using both log $z$\,--\,\,log $T$ and log $L$\,--\,\,log $T$ 90\% PEH bands.

\section{Data extraction methodology}
\label{section_data_extraction}
To extract PEH data, we follow the \emph{FromMax} method used by \citet{howell_2013} to untangle different populations of the \emph{Swift} GRB sample. This invoked the \emph{temporal cosmological principle}: for time scales that are short compared to the age of the Universe, there is nothing special about the time we switch on our detector. Therefore, the  time-series can be treated as closed loop, i.e. the start and endpoints of the time series can be joined and the start time set immediately after the brightest event. Successively brighter/closer events and their observation-times are then recorded to produce a PEH data set.

Employing this technique circumvents the possibility of a bright event occurring early in a time series; this would minimize the amount of output data as the next largest event would most probably occur near the end of the time series. Such a situation could be encountered through a detector with a high energy cutoff -- if a large number of events have energies around the threshold value a bias could be introduced.

Another feature of the FromMax method is that it establishes the total time duration of the PEH output to be equivalent to that of the total observation-time -- this ensures a well ordered data sample is produced with a consistent time signature. \citet{howell_2013} showed through statistical testing that the improved data set retains the statistical signature of the original.

To apply the procedure to a sample of $L(T)$ time-series data, we first define the brightest event by $L_*$ with an observation-time stamp $T_*$ and denote the time of the last, most recent occurring event, as $T_\mathrm{max}$. Treating the data as a closed loop we reorder the data starting from the first event after $ L_*$. The time stamps of the re-ordered data set $L^{\backprime}(T_{L}^{\backprime})$ are now defined as:
\vspace{-2mm}
\begin{equation}\label{peh_time}
  T^{\backprime}_{L}=\biggl \lbrace{ \begin{array}{lll}
                   T-T_* & T>T_* \\
                   \\
                   T+T_{\mathrm{max}}-T_* & T\leq T_*
                 \end{array}}
\end{equation}

In \citet{howell_2013}, to obtain a PEH data set the data was extracted from the first minimum $L_{\mathrm{min}}^{\backprime}=P_{i}^{\backprime} < L_{i+1}^{\backprime}$ -- this additional step was to minimise the effect of an early bright event. In this study we adopt a simpler approach; we take the first PEH event as the first above or equal to the median value of the sample $L_{\mathrm{med}}$. A PEH data set is then obtained by recording successively brighter events $(L_{i}^{\backprime},T_{L,i}^{\backprime})$ satisfying the condition $L_{i+1}^{\backprime}>L_{i}^{\backprime}$ for $L_{i}^{\backprime}\geq L_{\mathrm{med}}^{\backprime}$.

To determine the PEH data set in the redshift domain one applies similar principles, treating the data as a closed loop but re-ordering the data from the first event after the closest redshift event $Z_0$. The time-stamps for the re-ordered set $Z^{\backprime}(T^{\backprime}_{Z})$ are then given by:

\begin{equation}\label{peh_time}
  T_{Z}^{\backprime}=\biggl \lbrace{ \begin{array}{lll}
                   T-T_0 & T>T_0 \\
                   \\
                   T+T_{\mathrm{max}}-T_0 & T\leq T_0
                 \end{array}}
\end{equation}
% After again normalising the data set, $t_{z} = t_{z,i}/t_{z,1}$
\noindent A PEH data set is obtained by extracting data from the first event equal to or less than the median value of the distribution $Z_{\mathrm{med}}$, recording successively closer events $(Z_{i}^{\backprime},T_{Z,i}^{\backprime})$ satisfying the condition $Z_{i+1}^{\backprime}<Z{i}^{\backprime}$ for $Z_{i}^{\backprime}\geq Z_{\mathrm{med}}^{\backprime}$.

\section{Constraining the rate density of \emph{Swift} LGRBs}
\label{section_rate_results}
\begin{figure}
        \includegraphics[scale = 0.52,bbllx = 0cm,bblly = 0.3cm, bburx = 13.2cm, bbury = 11cm,origin=lr]{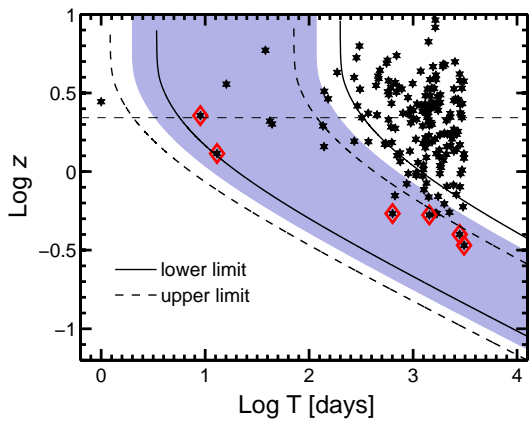}\\
  \caption{The log $z(T)$ distribution of \emph{Swift} LGRBs up to August 2013 using redshifts obtained through absorbtion spectroscopy or photometrically -- PEH data is indicated by the red diamonds. The median value for the distribution is shown by the horizontal dashed line. log $z$--log $T$ curves are used to support our rate estimates of $\rho_{0}=0.48^{+0.38}_{-0.24}$ Gpc$^{-3}$yr$^{-1}$ obtained through a $V_{\mathrm{max}}$ analysis. The shaded region represents the best estimate and thin dashed and solid curves show the lower and upper limits respectively. In each case the upper log $z$--log $T$ curve represents the 95\% confidence threshold of at least 1 occurrence within an observation time $T$. The lower log $z$--log $T$ curve indicates the 95\% confidence threshold of no occurrence occurring within $T$. We ignore the first PEH event in our analysis which occurred above the median threshold.}
  \label{Swift_logZlogT_fit}
\end{figure}

\subsection{$V_{\mathrm{max}}$ analysis of the LGRB sample}
\label{section_vmax}
As an initial estimate of $\rho_{0}$, the intrinsic beam uncorrected rate of LGRBs, we use the 9 LGRBs from the \emph{Swift} sample which have been recorded within a volume encompassed by $z =0.6$ ( $\sim$ 3.7 Gpc). Extending \citet{GuettaDellaValle_2007,CowardHowellPiran_2012,howell_2013} one can determine the rate through the maximum detection volumes of the sample:

\begin{equation}\label{eq_vmax}
    \rho_{0} = \sum_{i}^{\in z_{i} \leq 0.6} \frac{1}{V_{\rm{max},i}} \frac{1}{T} \frac{1}{\Omega} \frac{1}{\eta_{z,i}} \frac{1}{\eta_{P,i}}
\end{equation}

\noindent Here, $V_{\mathrm{max}}$ is the maximum volume out to which each burst could be detected, $T$ is the maximum observation-time for the sample, $\Omega$ is the sky coverage (1.33/4$\pi$).

The $V_{\mathrm{max}}$ method is highly sensitive to the value of the detector threshold -- this poses a problem when considering detectors with highly complex triggering mechanisms such as \emph{Swift}. To take into account the triggering efficiency for each individual burst, we first set a maximum flux limit of the detector by $6\times
10^{-9}\, \mathrm{erg}\, \mathrm{sec}^{-1}$, the minimum value for over 99\% of the \emph{Swift} LGRB sample, then weight each burst using the efficiency functions $\eta_{P}$ and $\eta_{z}$ from  sections \ref{section_swift_detection efficiency}.

\begin{table}
  \centering
  \begin{tabular}{ccccc}
\hline
\hline
GRB  & $z$ &    Peak Flux & $\eta_{P}$ & $\eta_{z}$  \\
     &     &    $ (10^{-6} \mathrm{erg}\, \mathrm{sec}^{-1}\, \mathrm{cm}^{-2})$   &  &  \\
\hline
\hline
060729    &  0.54 &0.074 & 0.22 &  0.73\\
081007    &  0.53 &0.18 & 0.42 &  0.73\\
090424    &  0.54 &5.4 & 0.95 &  0.73\\
090618    &  0.54 &2.6 & 0.91 &  0.73\\
%100418A   &  0.62 &0.05 & 0.15 &  0.69\\
101219B   &  0.55 &0.099 & 0.28 &  0.73\\
120714B   &  0.4 &0.017 & 0.045 &  0.79\\
130215A   &  0.6 &0.12 & 0.33 &  0.7\\
130427A   &  0.34 &30 & 0.99 &  0.82\\
\hline
\hline
\end{tabular}
  \caption{The data used to determine the rate of LGRBs from the \emph{Swift} sample using the $V_{\mathrm{max}}$ method.}
  \label{table_vmax_data}
\end{table}

Table \ref{table_vmax_data} outlines the values of redshift and peak flux for each of the bursts within $z=0.6$ and their relative scaling values, $\eta_{P}$ and $\eta_{z}$. We obtain a rate estimates of $\rho_{0}=0.48^{+0.38}_{-0.24}$ Gpc$^{-3}$yr$^{-1}$ where the errors are the 95\% Poisson confidence limits \citep{Gehrels_1986}.

\begin{figure}
    \includegraphics[scale = 0.51,bbllx = 0cm,bblly = 0.3cm, bburx = 13.2cm, bbury = 11cm,origin=lr]{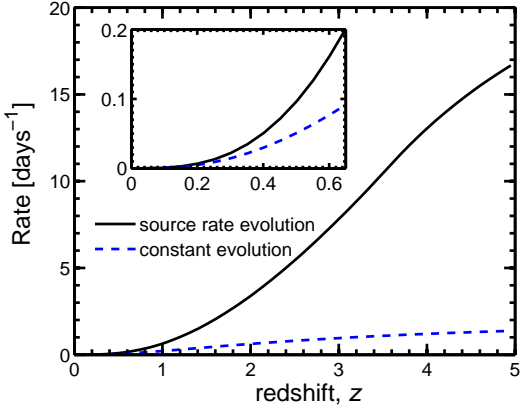}\\
  \caption{A comparison of the event rate of cosmic events using two different scenarions: an evolution based on LGRBs using equation \ref{eq_sfr}; a constant cosmic evolution. We see that within $z=0.6$ the two curves differ by a factor 2. }
  \label{fig_vmax_volume}
\end{figure}

We note that for this calculation will have ignored evolution effects within $z=0.6$. One can estimate the magnitude of any bias from this assumption by estimating a cosmic event rate:
\begin{equation}
\mathrm{d}R = \frac{\mathrm{d}V}{\mathrm{d}z}\frac{R_{\mathrm{GRB}}(z)}{1+z} \mathrm{d}z \,.
\label{eq_drdz}
\end{equation}
\noindent for two scenarios: a source rate evolution given by equation \ref{eq_sfr}; a constant evolution $R_{\mathrm{GRB}}(z)=1$. As shown by Figure \ref{fig_vmax_volume}, out to $z=0.6$ the two curves differ by around a factor of 2. Therefore, in our later calculations we will examine the LF across a large enough range around the above estimate to allow for any possible offset.

\subsection{The log\,$z$--\,log $T$ distribution}
\label{section_logZlogT_dist}
To test the validity of the $V_{\mathrm{max}}$ method we produce log $z$--log $T$ curves corresponding to the estimated rate $\rho_{0}=0.48^{+0.38}_{-0.24}$. Figure \ref{Swift_logZlogT_fit} shows the log $z$--log $T$ curves along with the log $z(T)$ distribution of \emph{Swift} LGRBs -- $z(T)$ PEH data is indicated by red diamonds. The median value for the distribution is shown by the horizontal dashed line. We note that we have excluded the first occurring PEH event in this analysis -- GRB 090424 (\emph{z} = 2.78). This event occurred above the median threshold and by our data extraction criteria discussed in section \ref{section_data_extraction} - the rare events from the tail of the distribution occur well within this value.

The log $z$--log $T$ curves are constructed using an observed rate, obtained by scaling $\rho_{0}$ by a value of 0.14 which represents the triggering efficiency of \emph{Swift} \citep{Lien_2014ApJ} and by the mean value of $\eta_{z}$ for our $V_{\mathrm{max}}$ sample which within $z=0.6$ equals 0.75.

The shaded region represents the best estimate and the thin dashed and solid curves show the lower and upper limits of $\rho_{0}$ respectively. In each case the upper log $z$--log $T$ curve represents the 95\% confidence threshold of at least 1 occurrence within an observation time $T$. The lower log $z$--log $T$ curve indicates the 95\% confidence threshold of no occurrence occurring within $T$. The PEH data is clearly constrained within the 90\% threshold defined by the two sets curves supporting the estimate of $\rho_{0}$ obtained through the $V_{\mathrm{max}}$ method.

\section{Constraining the LF of LGRBs}
\label{section_LF_results}
In this section we estimate the most compatible fitting parameters for the LGRB LF \{$\alpha$, $\beta$ and $L_{*}$ \} within the range $0.2 <\rho_{0} < 1$ Gpc$^{-3}$yr$^{-1}$. We will use both the \mbox{log\,$L$\,--\,log $T$} 90\% PEH bands and $\chi^{2}$ fitting of the log\,$N$--\,log\,$P$ distribution in an iterative procedure over $\rho_{0}$. We will firstly describe our procedure after which we will present our results.

\subsection{Constraining the LGRB LF through the \mbox{log\,$L$\,--\,log $T$} distribution}
To place constraints on the parameters of the LF before log\,$N$--\,log\,$P$ fitting is performed, the parameters \{$\rho_{0}$, $\alpha$, $\beta$ and $L_{*}$\} are required to constrain $L(T)$ data through log\,$L$\,--\,log $T$ 90\% PEH bands. The data is obtained by applying the methodology described in section \ref{section_data_extraction}. The log\,$L$\,--\,log $T$ data is given in Table \ref{table_LPEH_data}.

Using each set of \{$\rho_{0}$, $\alpha$, $\beta$ and $L_{*}$\} we construct log\,$L$\,--\,log $T$ 90\% PEH bands. A measure of compatibility is obtained through the binomial maximum likelihood (BML) \citep[BML;][]{UDD} estimate for obtaining data within the 90\% bands  -- a larger value signifies a good fit to the data; in this procedure a BML estimate of 88\% (1 failure in 8) will indicate a compatible set of parameters.

\begin{table*}
  \centering
  \begin{tabular}{ccccc}
\hline
\hline
GRB  & $T_{\mathrm{obs}}$ &  Peak Luminosity   &     Peak Flux                             & $z$  \\
     &     (days)               &    $ (10^{53} \mathrm{erg}\, \mathrm{sec}^{-1}\,)$   & $ (10^{-7} \mathrm{erg}\, \mathrm{sec}^{-1}\, \mathrm{cm}^{-2})$ &   \\
\hline
\hline
130514A &4.00 &0.37 &1.60 &3.60 \\
130606A &26.00 &3.09 &1.85 &5.91 \\
050401 &140.00 &4.92 &9.73 &2.90 \\
061007 &691.00 &5.73 &15.52 &1.26 \\
071020 &1069.00 &6.67 &9.02 &2.15 \\
080607 &1301.00 &17.97 &17.63 &3.04 \\
080721 &1345.00 &25.43 &18.89 &2.59 \\
130505A &3088.00 &30.29 &18.93 &2.27 \\
\hline
\hline
\end{tabular}
  \caption{The $L(T)$ data sample used to constrain the LF parameters.}
  \label{table_LPEH_data}
\end{table*}

\subsection{The log N-log P brightness distribution for Swift GRBs}
\label{subsection_logNlogP_results}
To perform $\chi^{2}$ minimisation of the log\,$N$--\,log\,$P$  brightness distribution of \emph{Swift} LGRBs we use 15-150 keV band peak energy flux data from Butler's online catalogue of \emph{Swift} BAT Integrated Spectral Parameters described in section \ref{section_LGRB_data_sample}. From equation \ref{dN} one can define a differential log\,$N$--\,log\,$P$ relation \citep{kommers_2000,pm01,salvaterra_07,Campisi2010,howell_2013} which is the observed rate of bursts within a peak flux interval ($P_{1}$, $P_{2}$)\footnote{We note that \mbox{log\,$(L;z)$--\,log $T$.} relations are derived from integral distributions. As bright objects will contribute to counts at all values of $P$ in an integral distribution, we fit peak flux data using a differential distribution in which the number of sources are independent at each interval of $P + \mathrm{d}P$}:
\vspace{-0.5mm}
\begin{equation}\label{eq_differential_peak_flux}
    \hspace{-1.0mm}
    \dot{N}(P_{1}\hspace{-0.5mm}\leq \hspace{-1.0mm}P\hspace{-1.0mm}< \hspace{-0.5mm}P_{2}) = \hspace{-1mm}\frac{\Omega \, \eta_{P}(P)}{4 \pi}\hspace{-2mm}\int\limits^{\hspace{3mm}\infty}_{\hspace{2mm}0} \frac{\mathrm{d}V (z) }{\mathrm{d}z } \frac{ R_{\mathrm{GRB}} }{( 1 + z)}\mathrm{d}z \hspace{-6.5mm} \int\limits^{\hspace{7mm}L(P_{2},z)}_{\hspace{6mm}L(P_{1},z)} \hspace{-4mm}\Phi(L) \mathrm{d}L\,,
\end{equation}

\noindent with ${L(P_{1,2},z)}$ obtained through equation \ref{eq_rest_frame_luminosity}. The peak flux triggering efficiency $\eta_{P}$ is given in equation \ref{swift_pf_efficiency_curve} (section \ref{peak_flux_efficiency}).

We bin peak flux data into logarithmically spaced intervals $\Delta P$ and ensure each bin contains at least 5 bursts \citep{PracStatsAst03}. Bursts per bin $\Delta N$ and their uncertainties $\pm \sqrt{\Delta N}$ are converted into burst rates $\Delta R$ by dividing by the live time of the search $\Delta T$ \citep{kommers_2000}. The peak flux intervals, number of bursts and burst rates data used for the fit is given in Table \ref{table_lognlogp_data}.

The goodness of fits are indicated through values of minimum $\chi^{2}$ per degree of freedom, $\chi^{2}/dof$, and $P_{\chi^{2}}$. The latter parameter is the probability of obtaining a $\chi^{2}$ value equal to or greater than $\chi^{2}$  given the data is drawn from the model using the best-fit parameters.

\subsection{Parameter Search}
For each value of $ 0.2 \leq \rho_{0}\leq 1.0$ we iterate through a range of LF parameters \{$L_{*}$, $\alpha$ and $\beta$\} to construct log\,$L$\,--\,log $T$ 90\% PEH bands. If the BML estimate for obtaining data within the 90\% bands is $\geq$ 88\% (1 failure in 8) we perform $\chi^{2}$ minimisation of the log\,$N$--\,log\,$P$ and store the value of $\chi^{2}$. We continue until a maximum $\chi^{2}$ is obtained.
\begin{table}
  \centering
  \begin{tabular}{cccc}
\hline
      P1  & P2  &     $\dot{N}$   &   $\Delta \dot{N}/ \Delta P$  \\
      $ \mathrm{erg}\, \mathrm{s}^{-1}\, \mathrm{cm}^{-2}$   & $ \mathrm{erg}\, \mathrm{s}^{-1}\, \mathrm{cm}^{-2}$ &  $\mathrm{yr}^{-1}$     &  $\mathrm{yr}^{-1}\, \mathrm{erg}\, \mathrm{s}\,\, \mathrm{cm}^{2}$ \\
      $ \times 10^{-7}$   & $\times 10^{-7}$ &       &  $\times 10^{5}$ \\
\hline
0.062  &  0.10 & 0.94 &  232\\
0.10  &  0.17 & 3.10 &  455\\
0.17  &  0.28 & 7.90 &  707\\
0.28  &  0.47 & 9.20 &  497\\
0.47  &  0.77 & 11.00 &  369\\
0.77  &  1.30 & 11.00 &  225\\
1.30  &  2.10 & 12.00 &  144\\
2.10  &  3.50 & 7.20 &  51.70\\
3.50  &  5.80 & 3.60 &  15.90\\
5.80  &  9.60 & 4.00 &  10.50\\
9.60  &  16.00 & 1.80 &  2.80\\
16.00  &  26.00 & 1.50 &  1.46\\
26.00  &  44.00 & 0.47 &  0.27\\
44.00  &  330.00 & 0.71 &  0.025\\
\hline
\end{tabular}
  \caption{The data used to fit the differential peak flux distribution of the \emph{Swift} long GRB sample. The data is obtained within the energy range 15-150 keV.}
  \label{table_lognlogp_data}
\end{table}

\subsection{Results}
Table \ref{table_fits_lognlogp_bpl} shows global LGRB parameters that satisfied the conditions of the parameter search (indicated by a $\checkmark$). We also show a few values each side of this range obtained by relaxing the selection criteria (indicated by a \text{\sffamily X}).

We found that parameters that passed both selection criteria were within the range $ 0.5 \leq \rho_{0}\leq 0.8$. For values outside this range, steeper values of $\beta$ were required to fit the log \emph{N}--log \emph{P} distribution. However, as discussed in section \ref{The logLlogT relation}, steep values of $\beta$ correspond with a lower proportion of bright bursts, resulting in log\,$L$\,--\,log $T$ curves that are too flat at the high $L$ end to constrain the brightest bursts.

Values of $\rho_{0}$ greater than 0.8\, $\mathrm{Gpc}^{-3}\, \mathrm{yr}^{-1}$ fared better than values below 0.5\, $\mathrm{Gpc}^{-3}\, \mathrm{yr}^{-1}$ which were ruled out by poorly constrained $L(T)$ PEH data (as shown, values of MLE=0.75 were obtained for 0.85-0.90\, $\mathrm{Gpc}^{-3}\, \mathrm{yr}^{-1}$).

\begin{table*}
  \centering
  \begin{tabular}{cccccccc}
\hline
\hline
 $\rho_{0}$  &        $L_{*}$ &   $\alpha$  &  $\beta$ & $\chi^{2}/dof$  & $P_{\chi^{2}}$ & BML & Pass/Fail\\
  $\mathrm{Gpc}^{-3}\, \mathrm{yr}^{-1}$  & $\times 10^{52} \mathrm{erg}\, \mathrm{sec}^{-1}$  &   &  &  &  & &\\
  \hline
$0.90$  &   $0.95 \pm0.80 $  &   $-1.15 \pm0.10$  &   $-2.66 \pm1.83$  &   1.538  &   13.41  &   0.75  & \text{\sffamily X} \\
$0.85$  &   $0.92 \pm0.79 $  &   $-1.10 \pm0.10$  &   $-2.66 \pm1.84$  &   1.525  &   14.05  &   0.75   & \text{\sffamily X}\\
\hline
$0.80$  &   $0.80 \pm0.43 $  &   $-0.95 \pm0.09$  &   $-2.59 \pm0.93$  &   1.239  &   33.04  & 0.88   &  $\checkmark$\\
$0.75$  &   $0.81 \pm0.56 $  &   $-0.93 \pm0.11$  &   $-2.60 \pm1.24$  &   1.242  &   32.74   & 0.88  &  $\checkmark$ \\
$0.70$  &   $0.70 \pm0.26 $  &   $-0.80 \pm0.10$  &   $-2.55 \pm0.67$  &   1.272  &   30.32   & 0.88  &  $\checkmark$ \\
$0.65$  &   $0.69 \pm0.30 $  &   $-0.74 \pm0.13$  &   $-2.53 \pm0.75$  &   1.366  &   23.26   & 0.88  &  $\checkmark$ \\
$0.60$  &   $0.56 \pm0.28 $  &   $-0.55 \pm0.20$  &   $-2.47 \pm0.75$  &   1.417  &   19.95   & 0.88  &  $\checkmark$ \\
$0.55$  &   $0.51 \pm0.26 $  &   $-0.34 \pm0.27$  &   $-2.44 \pm0.72$  &   1.474  &   16.63   & 0.88  &  $\checkmark$ \\
$0.50$  &   $0.48 \pm0.25 $  &   $-0.13 \pm0.49$  &   $-2.42 \pm0.71$  &   1.558  &   12.50   & 0.88  &  $\checkmark$ \\
\hline
$0.45$  &   $0.49 \pm0.47 $  &   $-0.03 \pm0.91$  &   $-2.52 \pm1.44$  &   1.917  &   2.94  &   0.88   & \text{\sffamily X}\\
$0.45$  &   $0.64 \pm0.14 $  &   $-0.02 \pm0.26$  &   $-2.99 \pm0.99$  &   1.165  &   39.55  &   0.13  & \text{\sffamily X}\\
\end{tabular}
  \caption{The results of a LF parameter search for sets of \{$L_{*}$, $\alpha$ and $\beta$\} that simultaneously satisfy a fit to the log $N$--log $P$ distribution (at a $P_{\chi^{2}} \geq 5\% level)$ and can constrain log $L$--log $T$ data, as indicated by a binomial maximum likelihood (BML) estimate for data within the 90\% PEH bands of 88\% (1 failure in 8). The log $N$--log $P$ goodness of fits for given by the the minimum $\chi^{2}$ per degree of freedom, $\chi^{2}/dof$, and $P_{\chi^{2}}$ -- the probability of obtaining a $\chi^{2}$ equal to or greater than $\chi^{2}$  given the data is drawn from the model using the best-fit parameters. Parameter tests that pass both criteria are indicated with a tick. Values above $\rho_{0}=0.8 \mathrm{Gpc}^{-3}\, \mathrm{yr}^{-1}$ were not able to constrain the PEH data at the BML=88\% level. For illustration we show two sets of parameters corresponding to $\rho_{0}=0.45 \mathrm{Gpc}^{-3}\, \mathrm{yr}^{-1}$ that both failed - the corresponding curves are given in Figure \ref{figure_logLlogT_extreme} and are described in the section \ref{section_LF_results}.}
  \label{table_fits_lognlogp_bpl}
\end{table*}

\begin{figure}
    \includegraphics[scale = 0.47,bbllx = 0cm,bblly = 2cm, bburx = 13.2cm, bbury = 12cm,origin=lr]{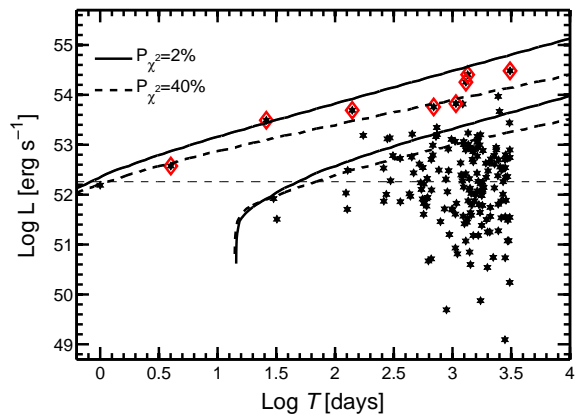}\\
  \caption{The log $L(T)$ distribution of \emph{Swift} LGRBs with PEH data shown as red diamonds. The median value for the distribution is shown by the horizontal dashed line. To illustrate how the parameter space of the LF can be further interrogated by complementing a log $N$--log $P$ fit with the log $L$--log $T$ method we show two sets of curves: each constructed with different LF parameters but both using the same rate of 0.45 $\mathrm{Gpc}^{-3}\, \mathrm{yr}^{-1}$ (see Table 4 for parameter values). \textbf{(Solid lines:)} The 90\% PEH bands constrain the PEH data at the MLE of 88\% level(1 fail in 8); however, the corresponding log $N$--log $P$ fits for the same set of parameters produced $P_{\chi^{2}}< 2\%$ \textbf{(Dashed lines:)} The PEH data is poorly constrained with this set of LF parameters (the MLE is 13\% - 1 fail in 8); however, the corresponding log $N$--log $P$ fits produced a good fit with $P_{\chi^{2}}=40\%$.
  }
  \label{figure_logLlogT_extreme}
\end{figure}

For clarity, Figure \ref{figure_logLlogT_extreme} shows an example of an incompatible parameter set for the case $\rho_{0}=0.45$. For this value, no parameters could be found that satisfied both criteria. For the PEH data to be constrained at the MLE=0.88 level, the best log \emph{N}--log \emph{P} fit gave only $\chi^{2}=2\%$ (solid line) and therefore failed the test. The best log \emph{N}--log \emph{P} fit (dashed line) produced a value of $\chi^{2}=40\%$. However, this parameter only managed to constrain 1 of the PEH data. No parameters could be found for $\rho_{0}=0.45$ that satisfied both criteria and therefore this value failed the test.

The compatible parameter sets (in the range 0.5--0.8 \,$\mathrm{Gpc}^{-3}\, \mathrm{yr}^{-1}$ ) are within the range of rates estimates calculated in section \ref{section_rate_results} --  however, a trend towards higher values of this range is evident. This offset is consistent with the factor of 2 discussed in section \ref{section_rate_results} and most likely results from neglecting source rate evolution within $z=0.6$ for the $V_{\mathrm{max}}$ calculation.

Rates of between 0.7 -- 0.8 $\mathrm{Gpc}^{-3}\, \mathrm{yr}^{-1}$ yield both well fitting log \emph{N}--log \emph{P} models ($\chi^{2}/dof\sim1.3$ and $P_{\chi^{2}}=30-33\%$) and well constrained log\,$L$\,--\,log $T$ data (MLE=0.83 and  $\mathcal{P}_{\mathrm{LT}}=3\%$). Within this range we take the set of parameters associated with 0.8 $\mathrm{Gpc}^{-3}\, \mathrm{yr}^{-1}$ (with the highest $P_{\chi^{2}}$ value) as the best estimate.

\begin{figure}
    \includegraphics[scale = 0.55,bbllx = 0cm,bblly = 0cm, bburx = 13.2cm, bbury = 12cm,origin=lr]{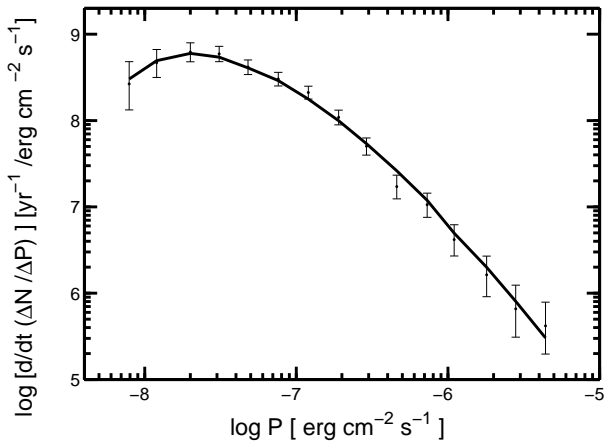}\\
  \caption{The log $N$\,--\,log $P$ distribution of \emph{Swift} LGRBs is fitted using using model parameters corresponding to the rate $\rho_{0}=0.8 \mathrm{Gpc}^{-3}\, \mathrm{yr}^{-1}$.}
  \label{Swift_logNlogP_best}
\end{figure}

Figure \ref{Swift_logNlogP_best} shows the fit to the log \emph{N}--log \emph{P} for this best set of parameters. The values of $\rho_{0}$ we obtain are in support of recent estimates of \citet{Butler_2010}, \citet{Wanderman2010} and L14. In particular, the use of the peak flux efficiency function based on the work of L14 support this study. Our results favour a steep value of $\beta$ as suggested by \citet{Butler_2010} and L14.

Compatible values of $\alpha$ shown in Table \ref{table_fits_lognlogp_bpl} are in agreement with the work of \citet{Zitouni2008MNRAS} who have shown that values of $-0.9 < \alpha < -0.4$ are fully consistent with the prediction of an internal shock model for $\gamma$-ray emission \citep{ReesMeszaros1994ApJ}. We note however that these values are slightly steeper than those obtained by L14 - this could be the result of a slight degeneracy between this parameter and $\rho_{0}$ which can both produce a vertical displacement. This suggests a LF with less free parameters could be beneficial to place additional constraints on the range of $\rho_{0}$ and will be considered in a future study.

\section{conclusions}
\label{section_conclusions}
When using number count relations to probe the rate and LF dependence of LGRBs, one is confronted by a number of difficulties. Firstly, poor modeling of the complex triggering criteria of the \emph{Swift} instrument can lead to large biases and poor estimates in parameters. Secondly, a complex mixing of the LF and source rate evolution can lead to degeneracies. Thirdly, for redshift dependent relations, the effect of high redshift selection biases can be difficult to quantify.

To confront the first of these obstacles we have used the results of a recent comprehensive study of the \emph{Swift} instrument by L14 to produce an efficiency function for peak flux. This removes the considerable uncertainty encountered by using a single value of flux to represent the triggering sensitivity limit of the detector. An additional efficiency function has been constructed in redshift space, again based on the results of L14.

In their study, L14 determined the best range of global parameters able to reproduce the observed distribution, based on the sample of \citet{Sakamoto2011ApJS} which comprised of 414 bursts. Applying our alternative methodology to a larger sample of 644 LGRBs, our results are in good agreement.

To approach the degeneracy problem we used a suite of different methods. We have complemented more standard $V_{\mathrm{max}}$ and log $N$--log $P$ brightness distribution methods with two observation-time relations, log $(L;z)$\,--\,log $T$. To determine initial estimates of the intrinsic (beaming uncorrected) rate density of LGRBs we have used the $V_{\mathrm{max}}$ method. To confront the small sample used in this method we have weighted the peak flux and redshift data using the corresponding efficiency functions. To verify this initial estimate we have constrained the log $z$\,--\,log $T$ distribution.

To determine the most likely parameters of the LGRB LF we have complemented a standard log $N$--log $P$ brightness distribution with a new log $L$\,--\,log $T$ relation. We have used both these methods simultaneously in an iterative procedure around the range of rate values calculated using $V_{\mathrm{max}}$. Our results support an event rate density at the high end of recent estimates \citep[][,L14]{Butler_2010,Wanderman2010,Cao_2011,Kanaan_2013A&A} and assuming a broken power law model for the LF, a steep high end slope.

There are a number of advantages in using the observation time relations. The log $L$\,--\,log $T$ method exploits only the brightest proportion of bursts which are less likely to be below the detection threshold. We have shown that this relation is sensitive to the form of the LF, particularly the bright end. For the log $z$\,--\,log $T$ relation, as the method is dependent on only the rarest close events, high-$z$ biases do not effect the analysis.

We have selected a broken power law LF model for this study for comparison with L14 and other recent studies that choose this form. We note that there is at present no preferred form of the LF with a number of studies choosing a power law model with an exponential cut-off \citep{howell_2013,Wei_2013} or log-normal distribution in luminosity \citep{Shahmoradi_13}. One advantage of using one of the latter forms, which have one less free parameter than a broken power model, is to minimise the risk of degeneracy in the parameter estimations. Our results suggest a small degeneracy between the rate density and the low end slope of the chosen LF. Future studies should therefore extend the work presented here and adopt a number of functional forms. A Bayesian approach would enable constraints set by the two observation time relations to be fed into the analysis as priors. Another possibility would be to investigate the form of the rate evolution, particularly the high redshift end.

To avoid added complexity, this study has not considered an evolving LF. The log $L$\,--\,log $T$ method could be used to probe changes in different redshift intervals. A particular advantage of this method is that a log $L$\,--\,log $T$ distribution is dependent on only the brightest proportion of data; the requirement of an adequate amount of data in each redshift interval could therefore be circumvented. An evolving LF can be modeled by modifying the break luminosity as $L_{*}(1 + z)^{\delta}$ with the value of $\delta$ generally in the range 1--2. As illustrated in section \ref{The logLlogT relation}, an increase in $L_{*}$ would produce a vertical offset in the log $L$\,--\,log $T$ 90\% PEH bands. This merits further investigation with a range of LF models and will be presented in a forthcoming study.

\section*{Acknowledgments}
E. J. Howell acknowledges support from a UWA Research Fellowship. D.M. Coward is supported by an Australian Research Council Future Fellowship.The authors gratefully acknowledge L14 for kindly providing the data used in section 3 and for valuable discussions in regards to modelling the detection efficiency of \emph{Swift}. The authors also acknowledge the anonymous referee for a careful reading of the manuscript and for providing a number of comments and suggestions that have significantly improved the paper.

\appendix

\onecolumn
\newpage

\section{The \emph{Swift} LGRB data sample}

\setlength\LTcapwidth{7.0in}
\begin{longtable}{llccllllccll}
\caption{The data for the LGRB sample used in this paper. Peak energy flux data is taken from Butler's online catalogue of \emph{Swift} BAT Integrated Spectral Parameters (\protect\url{http://butler.lab.asu.edu/Swift/bat_spec_table.html}) which is an extension of the work presented in \citet{Butler2007ApJ,Butler_2010}. Redshift data is taken from the Jochen Greiner online catalogue (JG) of well localized GRBs\,(\protect\url{http://www.mpe.mpg.de/~jcg/grbgen.html}). \emph{Type} indicates if the redshift was determined through absorbtion (A), emission (E), or photometry (P) in order of preference. Isotropic peak luminosity data is determined using the peak flux and redshift samples and corrected using the spectral parameters presented in the Butler catalogue.}
\\
\hline
  GRB  & $T_{obs}$ &    Peak Flux $\,\times 10^{-7}$  & $ L\,\times10^{53}$ & $z$  & type &  GRB  & $T_{obs}$ &    Peak Flux $\times10^{-7}$ & $ L \,\times10^{53}$ & $z$  & type \\
         &      days               &  $  (\mathrm{erg}\, \mathrm{s}^{-1}\, \mathrm{cm}^{-2})$   & $  (\mathrm{erg}\, \mathrm{s}^{-1})\, $ &       &            &   &      days               &  $ ( \mathrm{erg}\, \mathrm{sec}^{-1}\, \mathrm{cm}^{-2})$   & $ (\mathrm{erg}\, \mathrm{s}^{-1})\, $ &       &  \\
\hline
\\
050126    &  43 &4.3601 & 6.1142 &  1.29&E&061110B   &  692 &3.6977 & 145.96 &  3.44&A\\
050223    &  70 &4.4748 & 0.3678 &  0.592&E&061121    &  703 &190.31 & 221.53 &  1.31&A\\
050315    &  92 &9.1043 & 10.799 &  1.95&A&061126    &  708 &96.31 & 160.82 &  1.16&P\\
050318    &  95 &23.695 & 5.0851 &  1.44&A&061222A   &  734 &101.49 & 80.928 &  2.09&E\\
050319    &  96 &7.8381 & 30.694 &  3.24&A&061222B   &  734 &7.5264 & 31.051 &  3.36&A\\
050401    &  108 &97.264 & 491.85 &  2.9&A&070103    &  750 &9.1485 & 21.672 &  2.62&E\\
050505    &  142 &13.551 & 153.79 &  4.27&A&070110    &  757 &3.8389 & 2.8745 &  2.35&A\\
050525A   &  162 &394.91 & 9.0548 &  0.606&E&070129    &  776 &2.4624 & 1.8664 &  2.34&E\\
050730    &  227 &3.3526 & 33.971 &  3.97&A&070208    &  785 &4.2954 & 0.43081 &  1.17&E\\
050801    &  228 &11.134 & 7.3888 &  1.56&P&070306    &  813 &28.624 & 22.667 &  1.5&E\\
050814    &  241 &2.668 & 10.104 &  5.3&P&070318    &  825 &12.813 & 5.1693 &  0.836&A\\
050819    &  246 &2.248 & 2.3008 &  2.5&E&070411    &  848 &5.9716 & 20.434 &  2.95&A\\
050820A   &  247 &16.84 & 131.13 &  2.61&A&070419A   &  856 &0.63617 & 0.04915 &  0.97&A\\
050826    &  253 &2.5777 & 0.20559 &  0.297&E&070419B   &  856 &9.463 & 5.6712 &  1.96&E\\
050904    &  261 &4.0053 & 137.06 &  6.29&A&070506    &  873 &7.867 & 11.406 &  2.31&A\\
050908    &  265 &4.3194 & 18.98 &  3.34&A&070521    &  888 &61.971 & 82.317 &  1.35&P\\
050915A   &  272 &8.6173 & 43.837 &  2.53&E&070529    &  896 &6.5259 & 30.185 &  2.5&A\\
050922C   &  279 &75.827 & 60.587 &  2.2&A&070611    &  908 &4.2936 & 1.5831 &  2.04&A\\
051001    &  288 &1.8412 & 0 &  0&E&070612A   &  909 &10.263 & 1.315 &  0.617&E\\
051006    &  293 &23.732 & 15.479 &  1.06&E&070721B   &  948 &9.5099 & 128.46 &  3.63&A\\
051016B   &  303 &9.0619 & 1.7475 &  0.936&E&070802    &  959 &2.5299 & 5.145 &  2.45&A\\
051109A   &  326 &32.69 & 53.294 &  2.35&A&070810A   &  967 &9.6646 & 4.9436 &  2.17&A\\
051111    &  328 &21.247 & 7.1898 &  1.55&A&071003    &  1020 &55.702 & 117.12 &  1.6&A\\
051117B   &  334 &4.7448 & 0.25098 &  0.481&E&071010A   &  1027 &3.887 & 0.85028 &  0.98&A\\
060115    &  397 &5.6315 & 10.369 &  3.53&A&071010B   &  1027 &46.667 & 4.0776 &  0.947&A\\
060124    &  406 &5.894 & 9.268 &  2.3&A&071020    &  1037 &90.223 & 666.53 &  2.15&A\\
060202    &  414 &2.1264 & 0.33772 &  0.783&E&071031    &  1048 &1.8327 & 6.7129 &  2.69&A\\
060206    &  418 &21.136 & 56.526 &  4.05&A&071117    &  1064 &98.665 & 20.038 &  1.33&E\\
060210    &  422 &19.124 & 163.05 &  3.91&A&071122    &  1069 &1.2374 & 0.64926 &  1.14&A\\
060223A   &  435 &9.1729 & 27.322 &  4.41&A&080129    &  1141 &2.5635 & 43.344 &  4.35&A\\
060418    &  490 &46.459 & 41.085 &  1.49&A&080207    &  1149 &15.256 & 8.207 &  2.09&E\\
060502A   &  504 &12.83 & 17.357 &  1.51&A&080210    &  1152 &9.7309 & 24.762 &  2.64&A\\
060510B   &  512 &2.6904 & 12.723 &  4.9&A&080310    &  1182 &5.5655 & 18.748 &  2.42&A\\
060512    &  514 &2.8502 & 1.6975 &  2.1&E&080319B   &  1191 &482.58 & 654.62 &  0.937&A\\
060522    &  524 &2.0467 & 8.1948 &  5.11&A&080319C   &  1191 &43.143 & 23.703 &  1.95&A\\
060526    &  528 &9.3424 & 34.828 &  3.22&A&080330    &  1202 &3.6413 & 3.7003 &  1.51&A\\
060602A   &  534 &2.995 & 1.5063 &  0.787&E&080411    &  1213 &302.92 & 101.07 &  1.03&A\\
060604    &  536 &2.3936 & 4.0413 &  2.14&A&080413A   &  1215 &41.274 & 115.45 &  2.43&A\\
060605    &  537 &3.038 & 6.6894 &  3.78&A&080413B   &  1215 &129.91 & 14.177 &  1.1&A\\
060607A   &  539 &10.496 & 16.369 &  3.08&A&080430    &  1232 &17.374 & 2.6128 &  0.767&A\\
060707    &  569 &5.7076 & 8.2473 &  3.42&A&080516    &  1248 &17.045 & 0 &  0&P\\
060708    &  570 &14.366 & 5.902 &  1.92&P&080520    &  1252 &7.0829 & 167.94 &  1.54&E\\
060714    &  576 &8.4294 & 21.027 &  2.71&A&080603B   &  1265 &26.168 & 26.113 &  2.69&A\\
060719    &  581 &13.721 & 8.7699 &  1.53&A&080604    &  1266 &1.8366 & 1.1776 &  1.42&A\\
060729    &  591 &7.3714 & 0.46839 &  0.54&A&080605    &  1267 &179.1 & 79.228 &  1.64&A\\
060814    &  606 &52.025 & 32.964 &  1.92&E&080607    &  1269 &176.31 & 1797 &  3.04&A\\
060904B   &  626 &14.841 & 0.51546 &  0.703&A&080707    &  1299 &5.9124 & 2.8064 &  1.23&A\\
060906    &  628 &9.7544 & 50.78 &  3.69&A&080710    &  1302 &5.1389 & 3.2807 &  0.845&A\\
060908    &  630 &27.11 & 12.379 &  1.88&A&080721    &  1313 &188.9 & 2543.4 &  2.59&A\\
060912A   &  634 &63.361 & 16.417 &  0.937&E&080804    &  1326 &21.101 & 172.07 &  2.2&A\\
\\
\hline
\\
  GRB  & $T_{obs}$ &    Peak Flux $\,\times 10^{-7}$  & $ L\,\times10^{53}$ & $z$  & type &  GRB  & $T_{obs}$ &    Peak Flux $\times10^{-7}$ & $ L \,\times10^{53}$ & $z$  & type \\
         &      days               &  $  (\mathrm{erg}\, \mathrm{s}^{-1}\, \mathrm{cm}^{-2})$   & $  (\mathrm{erg}\, \mathrm{s}^{-1})\, $ &       &            &   &      days               &  $ ( \mathrm{erg}\, \mathrm{sec}^{-1}\, \mathrm{cm}^{-2})$   & $ (\mathrm{erg}\, \mathrm{s}^{-1})\, $ &       &  \\
\hline
\\
060926    &  648 &5.8964 & 62.226 &  3.2&A&080805    &  1327 &7.1877 & 7.4703 &  1.5&A\\
060927    &  649 &21.104 & 97.552 &  5.47&A&080810    &  1332 &11.623 & 101 &  3.35&A\\
061007    &  659 &155.23 & 572.97 &  1.26&A&080905B   &  1357 &9.3947 & 20.905 &  2.37&A\\
061021    &  673 &56.262 & 5.1363 &  0.346&E&080906    &  1358 &6.309 & 12.328 &  2.1&P\\
061110A   &  692 &3.2936 & 0.71671 &  0.758&E&080913    &  1365 &10.086 & 88.376 &  6.7&P\\
080916A   &  1368 &19.385 & 0.70595 &  0.689&A&100901A   &  2083 &3.1965 & 2.8301 &  1.41&E\\
080928    &  1380 &14.11 & 13.439 &  1.69&A&100906A   &  2088 &46.539 & 51.529 &  1.73&A\\
081007    &  1389 &17.789 & 0.39764 &  0.529&A&101219B   &  2191 &9.9342 & 0.55298 &  0.55&A\\
081008    &  1390 &8.5353 & 4.3352 &  1.97&A&110106B   &  2213 &15.383 & 1.2481 &  0.618&E\\
081028A   &  1410 &2.7577 & 3.8604 &  3.04&A&110128A   &  2235 &8.1931 & 60.826 &  2.34&A\\
081029    &  1411 &1.1961 & 10.627 &  3.85&A&110205A   &  2242 &18.762 & 13.645 &  2.22&A\\
081118    &  1430 &2.3908 & 6.211 &  2.58&A&110213A   &  2250 &66.638 & 40.871 &  1.46&A\\
081121    &  1433 &76.119 & 82.713 &  2.51&A&110422A   &  2319 &262.23 & 88.465 &  1.77&A\\
081203A   &  1445 &20.13 & 50.856 &  2.05&A&110503A   &  2330 &251.23 & 79.392 &  1.61&A\\
081222    &  1464 &60.885 & 72.919 &  2.77&A&110715A   &  2402 &452.93 & 23.841 &  0.82&A\\
081228    &  1470 &10.466 & 45.666 &  3.4&P&110731A   &  2418 &109.89 & 932.61 &  2.83&A\\
081230    &  1472 &3.5155 & 1.5149 &  2&P&110801A   &  2418 &6.0566 & 6.4104 &  1.86&A\\
090102    &  1479 &47.019 & 80.165 &  1.55&A&110808A   &  2425 &2.8731 & 1.8848 &  1.35&A\\
090205    &  1512 &2.1256 & 7.5285 &  4.65&A&110818A   &  2435 &9.8227 & 49.945 &  3.36&A\\
090313    &  1550 &2.4326 & 10.09 &  3.38&A&111008A   &  2485 &44.67 & 464.99 &  4.99&A\\
090417B   &  1584 &0.83035 & 0.019994 &  0.345&E&111107A   &  2514 &7.8189 & 35.944 &  2.89&A\\
090418A   &  1585 &15.633 & 26.893 &  1.61&A&111228A   &  2565 &69.345 & 7.4006 &  0.714&A\\
090423    &  1590 &9.813 & 112.27 &  8.26&P&111229A   &  2566 &6.9487 & 3.4264 &  1.38&A\\
090424    &  1591 &537.25 & 11.888 &  0.544&A&120119A   &  2591 &81.026 & 31.273 &  1.73&A\\
090426    &  1593 &24.989 & 58.938 &  2.61&A&120326A   &  2658 &26.094 & 10.186 &  1.8&A\\
090429B   &  1596 &9.9905 & 147.59 &  9.2&P&120327A   &  2659 &29.094 & 33.475 &  2.81&A\\
090516A   &  1613 &14.017 & 115.26 &  4.11&A&120404A   &  2666 &6.2705 & 18.199 &  2.88&A\\
090519    &  1616 &3.7118 & 87.486 &  3.85&A&120712A   &  2764 &18.292 & 50.36 &  4.17&A\\
090529    &  1626 &1.8599 & 1.5234 &  2.63&A&120714B   &  2766 &1.6976 & 0.012394 &  0.398&A\\
090530    &  1627 &26.345 & 18.461 &  1.3&P&120722A   &  2774 &9.7553 & 1.9846 &  0.959&E\\
090618    &  1645 &262.75 & 6.3057 &  0.54&A&120724A   &  2776 &1.9522 & 0.54163 &  1.48&A\\
090715B   &  1672 &24.827 & 98.968 &  3&A&120729A   &  2781 &17.904 & 3.6171 &  0.8&A\\
090726    &  1683 &3.7688 & 4.3975 &  2.71&A&120802A   &  2785 &17.174 & 34.619 &  3.8&A\\
090809    &  1696 &5.6521 & 16.536 &  2.74&A&120811C   &  2794 &22.944 & 22.192 &  2.67&A\\
090812    &  1699 &31.342 & 158.47 &  2.45&A&120815A   &  2798 &14.508 & 62.77 &  2.36&A\\
090814A   &  1701 &2.4645 & 0.074828 &  0.696&A&120907A   &  2821 &25.85 & 7.7046 &  0.97&A\\
090926B   &  1743 &19.793 & 2.417 &  1.24&A&120922A   &  2833 &9.4884 & 0 &  0&P\\
090927    &  1744 &17.167 & 9.5402 &  1.37&A&121024A   &  2864 &14.547 & 54.525 &  2.3&A\\
091018    &  1765 &53.127 & 6.0966 &  0.971&A&121128A   &  2899 &89.95 & 53.333 &  2.2&A\\
091020    &  1767 &30.443 & 11.581 &  1.71&A&121201A   &  2902 &5.4684 & 22.953 &  3.38&A\\
091024    &  1771 &16.944 & 17.669 &  1.09&A&121211A   &  2913 &4.5872 & 1.1832 &  1.02&A\\
091029    &  1776 &10.697 & 12.401 &  2.75&A&121229A   &  2931 &1.2273 & 3.07 &  2.71&A\\
091109A   &  1786 &8.5831 & 76.505 &  3.08&A&130131B   &  2962 &15.375 & 93.937 &  2.54&E\\
091127    &  1804 &375.48 & 14.929 &  0.49&E&130215A   &  2976 &12.316 & 1.1957 &  0.597&A\\
091208B   &  1815 &132.43 & 13.99 &  1.06&A&130408A   &  3029 &95.099 & 274.31 &  3.76&A\\
100219A   &  1891 &2.9994 & 42.685 &  4.67&A&130418A   &  3039 &1.4654 & 0.17435 &  1.22&A\\
100302A   &  1904 &2.1454 & 22.19 &  4.81&A&130420A   &  3041 &17.751 & 3.5178 &  1.3&A\\
100316B   &  1918 &7.237 & 3.1676 &  1.18&A&130427A   &  3048 &2986 & 55.576 &  0.34&A\\
100418A   &  1950 &4.9611 & 0.3468 &  0.624&A&130427B   &  3048 &22.982 & 78.77 &  2.78&A\\
100425A   &  1957 &4.7339 & 1.7868 &  1.75&A&130505A   &  3056 &189.29 & 3029.1 &  2.27&A\\
100513A   &  1975 &2.9867 & 29.833 &  4.77&A&130511A   &  3060 &20.399 & 15.428 &  1.3&A\\
100621A   &  2013 &77.33 & 1.9036 &  0.542&E&130514A   &  3063 &15.997 & 37.403 &  3.6&P\\
100724A   &  2046 &13.849 & 1.7775 &  1.29&A&130606A   &  3085 &18.546 & 309.19 &  5.91&A\\
100728B   &  2050 &24.41 & 43.74 &  2.11&A&130612A   &  3091 &8.3886 & 3.2176 &  2.01&A\\
100814A   &  2066 &14.81 & 3.6671 &  1.44&A&  &   & &  &  &\\
100816A   &  2068 &100.15 & 5.5693 &  0.805&A&  &   & &  &  &\\
%\\
%\hline
%\\
%  GRB  & $T_{obs}$ &    Peak Flux $\,\times 10^{-7}$  & $ L\,\times10^{53}$ & $z$  & type &  GRB  & $T_{obs}$ &    Peak Flux $\times10^{-7}$ & $ L \,\times10^{53}$ & $z$  & type \\
%         &      days               &  $  (\mathrm{erg}\, \mathrm{s}^{-1}\, \mathrm{cm}^{-2})$   & $  (\mathrm{erg}\, \mathrm{s}^{-1})\, $ &       &            &   &      days               &  $ ( \mathrm{erg}\, \mathrm{sec}^{-1}\, \mathrm{cm}^{-2})$   & $ (\mathrm{erg}\, \mathrm{s}^{-1})\, $ &       &  \\
%\hline
%\\
%\hline
%\hline
 \label{table_lum_data}
\end{longtable}
\twocolumn

\end{document}